\documentclass[twocolumn,longbibliography]{revtex4-2}
\usepackage{graphicx}
\usepackage{epsf}
\usepackage{amsmath,amssymb}
\usepackage{color}
\newcommand{\bvec}{\boldsymbol}
\newcommand{\bra}[1]{\langle #1|}
\newcommand{\ket}[1]{| #1\rangle}
\newcommand{\braket}[1]{\langle #1\rangle}
\newcommand{\bratoket}[2]{\langle #1|#2\rangle}

\def\C#1#2{{}^{#1#2}\textrm{C}}
\def\O#1#2{{}^{#1#2}\textrm{O}}

\def\Be#1{{}^{#1}\textrm{Be}}

\usepackage{ulem}
\begin{document}

%¥usepackage[doublespacing]{setspace}

%\preprint{KUNS-XXXX}
\title{Collective model for cluster motion in $\Be8$, $\C12$, and 
$\O16$ systems
based on microscopic $2\alpha$, $3\alpha$, and $4\alpha$ models
%with complex Gaussian centers
}

\author{Yoshiko Kanada-En'yo}
%\email{yenyo@ruby.scphys.kyoto-u.ac.jp}
\affiliation{
Department of Physics, Kyoto University, Kyoto 606-8502, Japan
}

\author{Nobuo Hinohara}
%\email{hinohara@nucl.ph.tsukuba.ac.jp}
\affiliation{
Center for Computational Sciences, University of Tsukuba, Tsukuba, 305-8577, Japan
}
\affiliation{
Faculty of Pure and Applied Sciences, University of Tsukuba, Tsukuba, 305-8571, Japan
}
\date{\today}

\begin{abstract}

A microscopic $n\alpha$ cluster model was applied to $\Be8$, $\C12$, and $\O16$ systems
to investigate cluster motion in the ground state and radial excitation.
In the microscopic calculation of  $\C12$ and $\O16$
using the generator coordinate method with the coordinate $D$ of the $\alpha$-$\alpha$ distance, 
excited states were obtained as the large-amplitude mode built on the ground state. 
A collective model was constructed to describe the cluster motion of these states
by utilizing inputs from the microscopic cluster model such as the norm kernel and energy expectation values.
Furthermore, the cluster model was extended by introducing the imaginary part of the coordinate $D$
to incorporate the dynamical effects on the collective mass.
The collective wave function obtained with the collective model was found to be in reasonable agreement with the results of the 
generator coordinate method 
for energies, root-mean-square radii, and amplitude functions.
\end{abstract}

\maketitle

\section{Introduction}

In nuclear systems,
large-amplitude collective motion plays an important role in
various nuclear structure phenomena such as ground-state correlation, cluster excitation,
and shape mixing, as well as in dynamical processes, such as cluster decay and nuclear fission.
To describe large-amplitude motion along a collective path
in a microscopic framework,
the generator coordinate
method~(GCM)~\cite{Hill:1952jb,Griffin:1957zza}
is one of the widely used approaches, particularly, in the 
study of cluster phenomena~\cite{Brink:1966,Brink:1968ybn,Horiuchi:1970,uegaki1,Fujiwara-supp,Descouvemont:1987zzb,LIBERTHEINEMANN1980429}.
However, application of the GCM is still limited  to a few generator coordinates mainly because of
the high computational cost of 
superposing a number of basis wave functions along the collective path,
which requires non-diagonal elements of the microscopic Hamiltonian.
To reduce the computational cost,
semi-microscopic approaches, such as the orthogonality condition model \cite{Saito:1969zz} and
phenomenological potential models, have been widely used
in cluster physics.
In many studies, potentials along the collective path are phenomenologically
adjusted to fit existing data; however, they are not based on a
fundamental derivation.

To derive the collective Hamiltonian by incorporating microscopic effects,
various prescriptions have been proposed
and are under development. One of the key problems is
how to evaluate the collective mass in the kinetic term.
However, as discussed in Ref.~\cite{Reinhard87},
%P -G Reinhard and K Goeke 1987 Rep. Prog. Phys. 50 1
some prescriptions, such as the cranking mass and Gaussian overlap approximation~(GOA) 
mass~\cite{Brink:1968ybn,Reinhard87}
are known to be insufficient to quantitatively describe collective motion.  
The center-of-mass motion is a typical example that both prescriptions fails to describe.
%The GOA method is another approach that incorporates dynamical effects; however,
i%t is not sufficient to quantitatively describe collective motion.
As further microscopic approach to
deriving the collective path and collective Hamiltonian, 
the self-consistent collective coordinate method has been developed~\cite{Marumori:1980bu},
and an adiabatic version has been applied to large-amplitude motion, including
shape mixing and cluster phenomena~\cite{Matsuo:2000hy,Hinohara:2007tj,Wen:2016buw}; 
however, it requires solving coupled equations.

In this study, we propose a convenient derivation
of a collective model
which can approximately describe the cluster dynamics obtained by
the microscopic calculation of the GCM.
We adopt the $2\alpha$, $3\alpha$, and $4\alpha$ models 
for the $\Be8$, $\C12$, and $\O16$ systems, respectively.
To describe cluster motion in the ground state and radial excitation,
we use the Brink-Bloch cluster wave functions~\cite{Brink:1966} with the 
most symmetric $n\alpha$ configurations, namely, 
the dumbbell, equilateral triangle, and regular tetrahedron configurations of $2\alpha$, $3\alpha$, and $4\alpha$, respectively.
The $\alpha$-$\alpha$ distance is defined by parameter $D$, and the cluster motion along the coordinate $D$ is considered.
First, the ground and excited states are microscopically calculated by the GCM 
using the generator coordinate $D$, and the cluster motion in the obtained states is analyzed.
Then, a collective model for the one-dimensional motion along the coordinate $D$ is constructed
by utilizing inputs from the microscopic $n\alpha$ wave functions, such as
the norm kernel and energy expectation values to 
derive the collective Hamiltonian.
Moreover, to incorporate the dynamical effects on the collective mass, 
the microscopic cluster model is extended by introducing the imaginary part of the coordinate $D$.
The collective Hamiltonian is evaluated by comparing the results for
the energies, root-mean-square radii, and collective wave functions of the ground and excited states
with the microscopic results
obtained by the GCM.

This paper is organized as follows.
Sections \ref{sec:cluster} and \ref{sec:hamiltonian} describe
the microscopic $n\alpha$ model and the microscopic Hamiltonian, respectively.
The GCM results for the microscopic calculation are presented in Sec.~\ref{sec:GCMresults}, while
the framework and results of the collective model are described in
Sec.~\ref{sec:collective}. A summary is provided in Sec.~\ref{sec:summary}.
Appendix \ref{app:coordinates} presents a detailed derivation of the physical coordinates.

\section{microscopic $n\alpha$ cluster model} \label{sec:cluster}

\subsection{Wave functions of $n\alpha$ cluster system}
A basis $n\alpha$ wave function 
is expressed by the Brink-Bloch alpha-cluster wave function~\cite{Brink:1966,Brink:1970ufk} as
\begin{equation}
\Phi_{n\alpha}(\bvec{S}_1,\ldots,\bvec{S}_n)=n_0
{\cal A}\left\{\psi_\alpha(\bvec{S}_1)
\cdots \psi_\alpha(\bvec{S}_n) \right\},
\end{equation}
where ${\cal A}$ is the antisymmetrizer,
and $\psi_\alpha(\bvec{S}_m)$ is the $\alpha$-cluster wave function
\begin{align}
&\psi_\alpha(\bvec{S}_m)=\prod_{i\in \alpha_m}
\phi_{\bvec{S}_m}(\bvec{r}_i) \chi_i\tau_i, \nonumber\\
&\phi_{\bvec{S}_m}(\bvec{r}_i) = \left(\frac{2\nu}{\pi}\right)^{3/4}
\exp\bigl[-\nu({\bvec{r}_i}-\bvec{S}_m)^2\bigr]
\end{align}
with the nucleon-spin and -isospin functions
$\chi_i\tau_i$ selected as $p\uparrow$,
$p\downarrow$, $n\uparrow$, and
$n\downarrow$ for four nucleons $i=4(k-1)+1,\ldots,4(k-1)+4$.
The Gaussian width parameter $\nu$ is set to $0.235$ fm$^{-2}$ in the present calculation.

The parameter
$\bvec{S}_m$ is usually treated as a real variable and 
indicates the mean center position of the $m$th $\alpha$-cluster $(\alpha_m$) in the coordinate space before antisymmetrization
in the original cluster model. However, complex variables for the Gaussian-center parameters $\bvec{S}_m$ are used in an extended 
cluster models, as discussed in a later section.
%\subsection{Generator coordinate method of cluster model}

To describe the inter-cluster motion  in the GCM approach,
the basis $n\alpha$ wave functions are superposed
with respect to the generator coordinates
$\bvec{S}_m$ as
\begin{align}
&\Psi_\textrm{GCM}=\nonumber\\
&\int d\bvec{S}_1,\ldots,d\bvec{S}_n
f(\bvec{S}_1,\ldots,\bvec{S}_n) \Phi_{n\alpha}(\bvec{S}_1,\ldots,\bvec{S}_n),
\end{align}
where the coefficients $f(\bvec{S}_1,\ldots,\bvec{S}_n)$ are determined by solving
the Hill-Wheeler equation~\cite{Hill:1952jb}.

\subsection{$2\alpha$, $3\alpha$, and $4\alpha$ models for $\Be8$, $\C12$, and $\O16$ systems}

\subsubsection{Model space}
In the $2\alpha$ system for $\Be8$, we
define the
distance parameter $D=|\bvec{S}|$
by taking $\bvec{S}_1=-\bvec{S}_2=\bvec{S}/2$ and
consider the relative motion
between two $\alpha$s in one-dimensional
model space distance with $D$.

\begin{figure}[!h]
\includegraphics[width=6 cm]{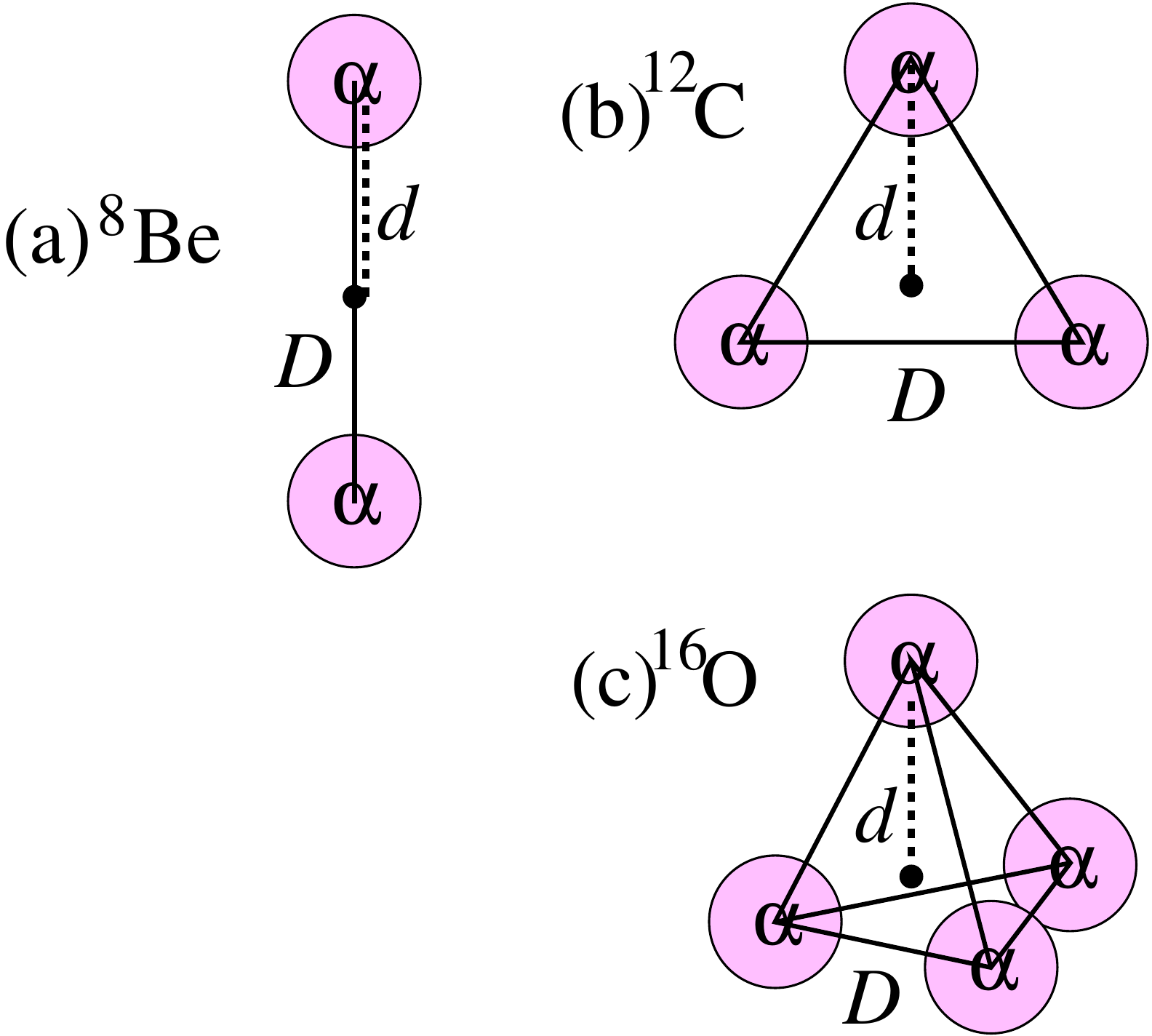}
\caption{Spatial configurations of the
$2\alpha$, $3\alpha$, and $4\alpha$ models for $\Be8$, $\C12$, and $\O16$
systems.
\label{fig:nalpha}}
\end{figure}

To describe the ground states and radial excitation
of the $3\alpha$ and $4\alpha$ systems
for $\C12$ and $\O16$,
we take highly symmetric configurations
by setting the $\alpha$-cluster positions
in the equilateral triangle and regular tetrahedron configurations 
as illustrated in Fig.~\ref{fig:nalpha}(b) and (c), respectively, and
we describe the radial motion of $\alpha$s 
using the GCM approach.
We define the $\alpha$-$\alpha$ parameter $D=|\bvec{S}_m-\bvec{S}_l|$ $(k\ne l)$.
Note that the radial distance $d=|\bvec{S}_m|$
measured from the origin is obtained by scaling $D$ as $d=D/2$,
$D/\sqrt{3}$, and
$\sqrt{3/8}D$, in the $2\alpha$, $3\alpha$, and
$4\alpha$ systems, respectively. The GCM calculation with the coordinate $D$
is equivalent to that with $d$. 

\subsubsection{GCM calculation}
In the GCM calculation, the parity-projected
wave functions $\Phi^\pm_{n\alpha}(D_j)$ at the mesh points of the coordinate $D_j$
are superposed as
\begin{align}
\Psi^\pm_{k}&=\sum_{j} f^\pm_k (D_j)
\Phi^\pm_{n\alpha} (D_j), \\
\Phi^\pm_{n\alpha} (D_j)&\equiv
\frac{ \hat{P}^\pm
\Phi_{n\alpha} (D_j)}
{\bratoket{\hat{P}^\pm \Phi_{n\alpha} (D_j)}
{\hat{P}^\pm \Phi_{n\alpha} (D_j)}^{1/2}},
\end{align}
where $\hat{P}^\pm$ is the parity-projection operator. 
Sixteen meshpoints $D_j=\{0.5 \textrm{~fm},\ldots,8.0\textrm{~fm}\}$ $(j=1,\ldots,16)$ with an interval of  $\Delta_D=0.5$~fm
are used in the present calculations.
The energy $E^\pm_{\textrm{GCM},k}$
and coefficients $f^\pm_k(D_j)$ for the $k$th parity($\pm$) state
are determined by diagonalization of the
norm and Hamiltonian matrices
\begin{align}
{\cal N}^\pm_{ij}&=\bratoket{\Phi^\pm_{n\alpha} (D_i)}{\Phi^\pm_{n\alpha} (D_j)},\\
{\cal H}^\pm_{ij}&=
\bra{\Phi^\pm_{n\alpha} (D_i)}\hat H \ket{\Phi^\pm_{n\alpha} (D_j)}.
\end{align}

For the GCM wave function $\Psi^\pm_{k}$ obtained for the $(\pm)_k$ states, we define
the amplitude function $G^\pm_{\textrm{GCM},k}(D_i)$ as
\begin{align}
G^\pm_{\textrm{GCM},k} (D_i)&\equiv \sum_{j} \left\{ {{\cal N}^{1/2}} \right\}_{ij} f^\pm_k (D_j),
\end{align}
where ${\cal N}^{1/2}$ is the square root of the $16\times 16$ matrix ${\cal N}^\pm_{ij}$ in the basis space
$D_i$ $(i=1,\cdots,16)$.
Since the norm kernel ${\cal N}^\pm_{ij}$ is defined in a finite-size subspace $D_i \leq 8.0$~fm of the full $D$ space,
$G^\pm_{\textrm{GCM},k} (D_i)$ sometimes exhibits oscillatory behavior
near the boundary $D_i\sim 8.0$~fm because of the finite-size effect; however, it is out our
region of  interest.
We also calculate the overlap function $G^{(2),\pm}_{\textrm{GCM},k}(D_i)$ of $\Psi^\pm_{k}$
with a basis wave function
$\Phi^\pm_{n\alpha} (D_i)$ at $D_i$ as
\begin{align}\label{eq:g2}
G^{(2),\pm}_{\textrm{GCM},k} (D_i)&\equiv \bratoket
{\Phi^\pm_{n\alpha} (D_i)}
{\Psi^\pm_{k}}
= \sum_{j}{\cal N}^\pm_{ij} f^\pm_k (D_j).
\end{align}
Note that
owing to the orthonormality $\bratoket{\Psi^\pm_{k}}{\Psi^\pm_{l}}=\delta_{kl}$,
the amplitude function 
$G^\pm_{\textrm{GCM},k} (D_i)$ satisfies the orthonormal condition
\begin{align}\label{eq:g-orthonormal}
& \sum_i G^{\pm *}_{\textrm{GCM},k} (D_i) G^\pm_{\textrm{GCM},l} (D_i)=\delta_{kl}, 
\end{align}
but the overlap function $G^{(2),\pm}_{\textrm{GCM},k} (D_i)$
does not.
Considering the transformation $\Delta_D \sum_{D_i}\to \int dD$, 
we redefine the amplitude function 
$g^\pm_{\textrm{GCM},l} (D) \equiv G^\pm_{\textrm{GCM},l} (D)/\sqrt{\Delta_D}$
to satisfy the standard normalization in the coordinate $D$ space as 
\begin{align} \label{eq:orthonormal-g}
&\int  g^{\pm *}_{\textrm{GCM},k} (D) g^\pm_{\textrm{GCM},l} (D) dD=\delta_{kl}.
\end{align}

%\subsubsection{Expectation values}
%The energy for the basis wave function at $D_i$ is given by
%the diagonal element $E(D_i)={\cal H}_ii}$.
%The expectation values such as the root-mean-square radius of the $k$th parity$(\pm)$ state
%are calculated for the corresponding
%operators such as $\hat H}$ and $\sum_i (\bvec{r}_i-\bvec{r}_G)^2$

%%%%%%%%%%%%%%%%%%%%%%%%%%%%%%
%%%%%%%%%%%%%%%%%%%%%%%%%

\section{microscopic Hamiltonian} \label{sec:hamiltonian}
The microscopic Hamiltonian $\hat H$ used in the present $n\alpha$ model 
consists of the single-nucleon kinetic energy, effective nucleon-nucleon($NN$) forces,
and $NN$ Coulomb force.
We use the effective central nuclear force given in a two-range Gaussian form
of the Volkov No.2 force~\cite{Volkov:1965zz} with the Majorana parameter $m=0.62$,
which has been used for cluster models in many studies.
The total center of mass (cm) motion can be exactly separated, and its kinetic energy is subtracted from the Hamiltonian.

\section{Results of microscopic calculation}\label{sec:GCMresults}

\subsection{Energies and radii}

%%%%%%%%%%%%%%%%%%%%%%%%%%%%%%
\begin{table}[!ptbh]
\caption{Energies and root-mean-square radii (rmsr) calculated with the microsopic $n\alpha$ model of
$\Be8$, $\C12$, and $\O16$. The values obtained by the generator coordinate method(GCM) and by single-basis calculation at
the minimum-energy distance $D_0$ are presented in the first and second columns, respectively. The $D_0$ values are displayed in parentheses.
For $\C12(+)$ and $\O16(+)$,
the values of two-basis diagonalization
for the small-amplitude calculation  are presented in the third column.
\label{tab:rmsr}
}
\begin{ruledtabular}
\begin{tabular}{lccccccc}
& GCM & single & small-amp. \\
&	$E$,	rmsr	&	$E$,	rmsr	($D_0$)	&	$E$,	rmsr	\\
&	[MeV], [fm]	&	[MeV], [fm]&	[MeV], [fm]	\\ \hline
$\Be8(+)_1$	&$	-46.3 ,~	2.67 $&$	-44.5 ,~	2.37 (3.2)	$&$ $\\
$\C12(+)_1$	&$	-74.9 ,~	2.35 $&$	-73.9 ,~	2.31 (2.2)	$&$	-73.8 ,~	2.30 $\\
$\C12(+)_2$	&$	-60.1 ,~	2.90 $&$ $&$	-52.8 ,~	2.37 $\\
$\C12(-)_1$	&$	-67.8 ,~	2.58 $&$	-66.7 ,~	2.54 (3.0)	$&$ $\\
$\O16(+)_1$	&$	-127.1 ,~	2.20 $&$	-126.7 ,~	2.19 (1.0)	$&$	-126.7 ,~	2.19 $\\
$\O16(+)_2$	&$	-102.6 ,~	2.46 $&$ $&$	-95.6 ,~	2.24 $\\
$\O16(-)_1$	&$	-113.7 ,~	2.34 $&$	-113.0 ,~	2.31 (1.7)	$&$ $\\
\end{tabular}
\end{ruledtabular}
\end{table}
%%%%%%%%%%%%%%%%%%%%%%%%%%%%

The positive- and negative-parity states of $\Be8$, $\C12$, and $\O16$
were calculated by the GCM of the $n\alpha$ model.
The  results obtained for the energies and root-mean-square radii
for the lowest $(\pm)_1$ and first excited states $(\pm)_2$ of each parity are presented in Table \ref{tab:rmsr}.
The calculated energy of the $\O16(+)_1$ state is in reasonable agreement with the experimental value
$E^\textrm{exp}(\textrm{gs})=-127.6$~MeV of the $\O16$ ground state,
whereas the calculated energies of the $\Be8(+)_1$ and $\C12(+)_1$ states are more than 10~MeV higher than
the experimental values $E^\textrm{exp}(\textrm{gs})=-56.50$~MeV and $-92.16$~MeV,
respectively.
These binding energy defects can be partially 
explained by the energy gain due to the total angular momentum projection for $\Be8$ and $\C12$,
and also by the spin-orbit attraction in $\C12$, neither of which is taken into account
in the present $n\alpha$ model.

The lowest negative-parity states $\C12(-)_1$ and $\O16(-)_1$ correspond to the
$K^\pi=3^-$ bands because of the point-group symmetry of the triangle and tetrahedron configurations,
respectively.
The calculated excitation energy $E_x$ of the $\C12(-)_1$ state, 7.1~MeV,  is in reasonable agreement 
with the experimental value $E^\textrm{exp}_x(3^-_1)=9.64$~MeV
of the band-head state, whereas that of $\O16(-)_1$, 13.4~MeV, estimates the experimental value
 $E^\textrm{exp}_x(3^-_1)=6.13$~MeV of the $\O16(3^-_1)$ state.

We also present the single-basis calculation results
at the minimum-energy distance $D_0$ optimized
for the diagonal elements of the Hamiltonian
$E^\pm(D)=\bra{\Phi^\pm_{n\alpha}(D)}\hat H\ket{\Phi^\pm_{n\alpha}(D)}$.
A comparison of the single-basis and GCM results indicates that the
$\C12(\pm)_1$ and $\O16(\pm)_1$ states can be approximately described by
the single configuration at $D_0$, whereas the $\Be8(+)_1$ state cannot be. This 
indicates that the $\alpha$-$\alpha$ motion in $\Be8(+)_1$ is not localized at a fixed distance, but
exhibits large fluctuation along the generator coordinate $D$, which significantly increases the binding energy gain and radius 
of $\Be8(+)_1$ in the GCM calculation.

The first excited positive-parity states, $\C12(+)_2$ and $\O16(+)_2$, are radial excitation
on the lowest states and have larger radii than the $\C12(+)_1$ and $\O16(+)_1$ states.
The $\C12(+)_2$ and $\O16(+)_2$ states are associated with monopole excitation but cannot be
assigned to the observed $0^+$ states in the $\C12$ and $\O16$ spectra. These theoretical radial excitation modes may
couple 
with other degrees of freedom, including $\alpha$-cluster motion and single-particle excitation, 
and may be fragmented into several cluster states and also partially contribute
to the giant monopole resonance.
For $\Be8$, no excited state is obtained as a bound-state solution.

\subsubsection{Potential energy curve and radial motion}

%%%%%%%%%%%%%%%%%%%%%%%%%%%%%%
\begin{figure*}[!htpb]
\includegraphics[width=18 cm]{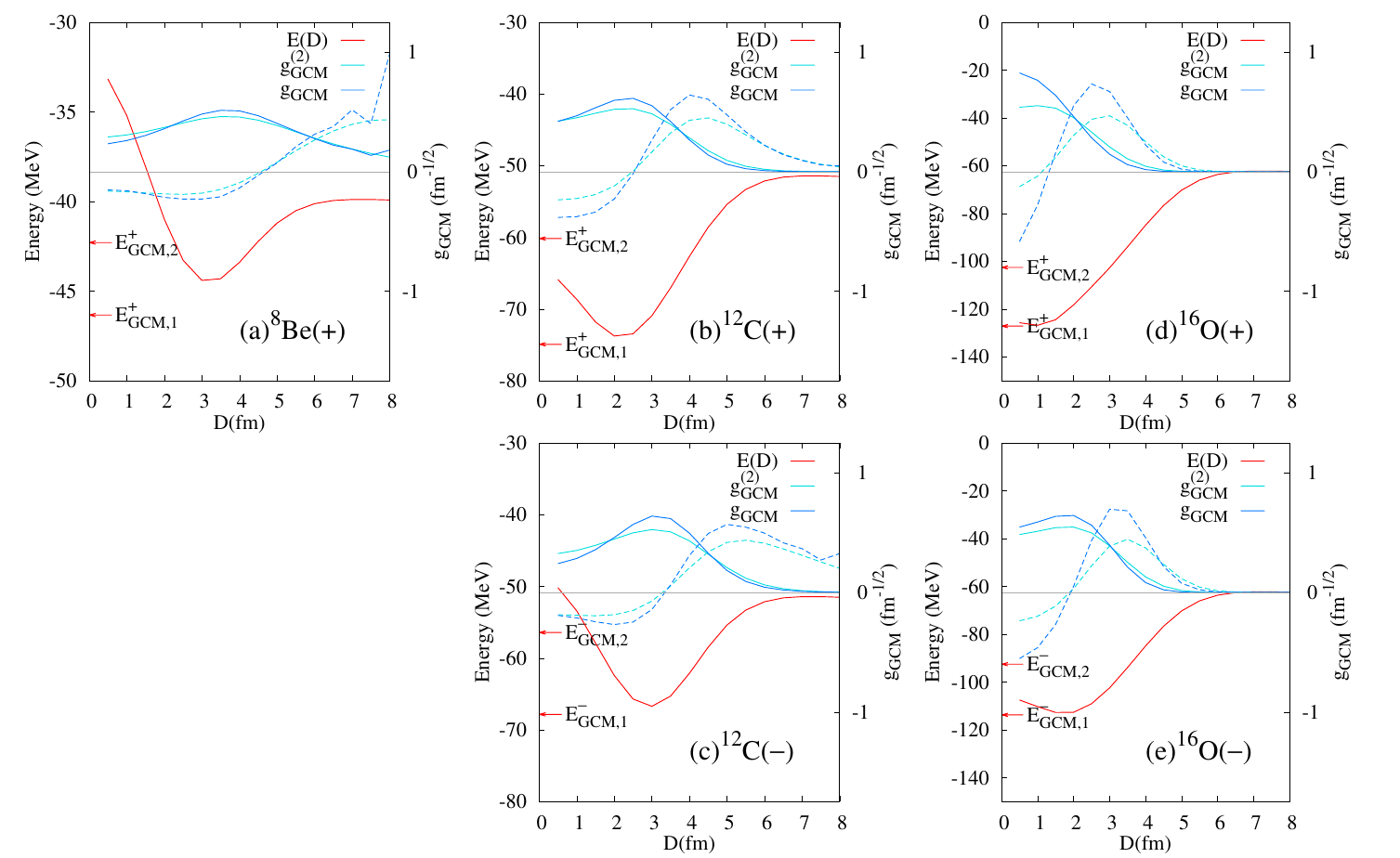}
\caption{GCM results of the energy curves
$E^\pm(D)$, GCM energies $E^\pm_{\textrm{GCM},k}$, and GCM amplitude functions $g^\pm_{\textrm{GCM},k}(D)$
for (a)$\Be8(+)$, (b)$\C12(+)$, (c)$\C12(-)$, (d)$\O16(+)$, and (e)$\O16(-)$.
For comparison, scaled values of the overlap function, 
$g^{(2),\pm}_{\textrm{GCM},k}(D)=c G^{(2),\pm}_{\textrm{GCM},k}(D)$, are also shown. 
Solid (dashed) lines indicate the amplitude and overlap functions of lower $(\pm)_1$ [higher $(\pm)_2$] states.
\label{fig:gcm}}
\end{figure*}
%%%%%%%%%%%%%%%%%%%%%%%%%

To discuss the radial motion along the generator coordinate $D$,
we present the potential energy curve $E^\pm(D)$, the GCM energies $E^\pm_{\textrm{GCM},k}$, and the GCM amplitude functions 
$g^\pm_{\textrm{GCM},k}(D)$ for the $(\pm)_1$ and $(\pm)_2$ states
in Fig.~\ref{fig:gcm}.

In $\Be8$, the energy curve displays effective repulsion in the interior region
because of the antisymmetrization effect and a shallow energy pocket at $D\sim 3$~fm.
The GCM amplitude $g^+_{\textrm{GCM},1}(D)$ is spread widely around the energy minimum.
As the number $n$ of $\alpha$ clusters increases in $\C12$ and $\O16$,
the energy pocket increases and shifts toward the interior region. Consequently,
the amplitude $g^+_{\textrm{GCM},1}$ of the $\C12(+)_1$ state is drawn inward, and that of the
$\O16(+)_1$ state concentrates around $D\sim 0$
to form a compact $4\alpha$ state, which approximately corresponds to
the $p$-shell closed state. For the excited positive-parity states $(+)_2$ of $\C12$ and $\O16$,
one can observe nodal behavior of $g^+_{\textrm{GCM},2}$
exhibiting vibration features of the radial ($D$) excitation built on the lowest states.

For the negative-parity states, the potential energy increases particularly
in the interior region, and the energy minima slightly shift outward.
Moreover, the amplitudes $g^-_{\textrm{GCM},k}$
shift somewhat outward, but they show qualitatively
similar features to the positive-parity states.
In other words, the amplitudes for the $(-)_1$ states concentrates
around the energy minima and those for the $(-)_2$ states exhibit the features of the 
vibrational excitation constructed on the $(-)_1$ states along $D$.

These results support the possible interpretation of the GCM amplitude $g^\pm_{\textrm{GCM},k}$
as ``collective wave functions'' in the collective coordinate $D$  for the $(\pm)_{1,2}$ states.
It should be noted that, although  $g^\pm_{\textrm{GCM},k}$ satisfies the 
orthonormal condition \eqref{eq:orthonormal-g}  for the $D$ integral,  
it does not satisfy the negative-parity 
boundary condition $g^-_\textrm{GCM}(0)=0$ naively expected from 
the transformation
$g^-_\textrm{GCM}(D)=-g^-_\textrm{GCM}(-D)$. 
It is indicated that the boundary condition at $D=0$ of the collective wave function is not trivial
because of the strong microscopic effects in the interior region close to $D=0$.

Figure~\ref{fig:gcm} also displays the overlap function $G^{(2),\pm}_{\textrm{GCM},k}(D)$
given in Eq.~\eqref{eq:g2}. To compare it with the amplitude function, 
we plot scaled values $g^{(2),\pm}_{\textrm{GCM},k}\equiv c G^{(2),\pm}_{\textrm{GCM},k}$
with a factor $c=\left(\frac{2 \nu 4(A-4)}{A\pi}\right)^{1/4}$,
as performed in
Ref.~\cite{Kanada-Enyo:2014mri} to evaluate inter-cluster wave functions from the norm overlap.
The $D$-dependence of the overlap function $g^{(2),\pm}_{\textrm{GCM},k}$ is qualitatively
consistent with that of the amplitude function $g^\pm_{\textrm{GCM},k}$
at least for the peak and node positions. This result indicates that
 $|g^{(2),\pm}_{\textrm{GCM},k}(D)|^2$ can be an alternative quantity for
the probability at $D$ as expected from the
physical meaning of the overlap. However, it should be kept in mind that neither 
$g^{(2),\pm}_{\textrm{GCM},k}(D)$ nor $G^{(2),\pm}_{\textrm{GCM},k}(D)$ 
satisfies the orthonormal condition, which may be a significant problem in
 associating them with a kind of ``collective wave function.''

\section{Collective model} \label{sec:collective}
The aim of this section is to construct
a potential model for the collective motion along the
parameter $D$ in $n\alpha$ systems.
The key question is
how to derive a collective Hamiltonian,
that can approximately describe the microscopic results obtained by the GCM.
%In particular, carefully
%interier region, in which dynamics is strongly influenced by antisymmetriationi effects.
For this purpose, we first describe 
the antisymmetrization effects on the coordinate space of
$D$ in the microscopic wave function,
and consider an extension of the real parameter $D$ to
a complex variable to discuss the dynamical effects on the cluster motion.
Then we propose a collective Hamiltonian for the cluster motion. 
Finally, we present the results obtained by solving the collective Hamiltonian, 
and compare them with the GCM results.

\subsection{Antisymmetrization effect on collective coordinate distance $D$}

\subsubsection{Physical meaning of $D$}

\begin{figure}[!h]
\includegraphics[width=8 cm]{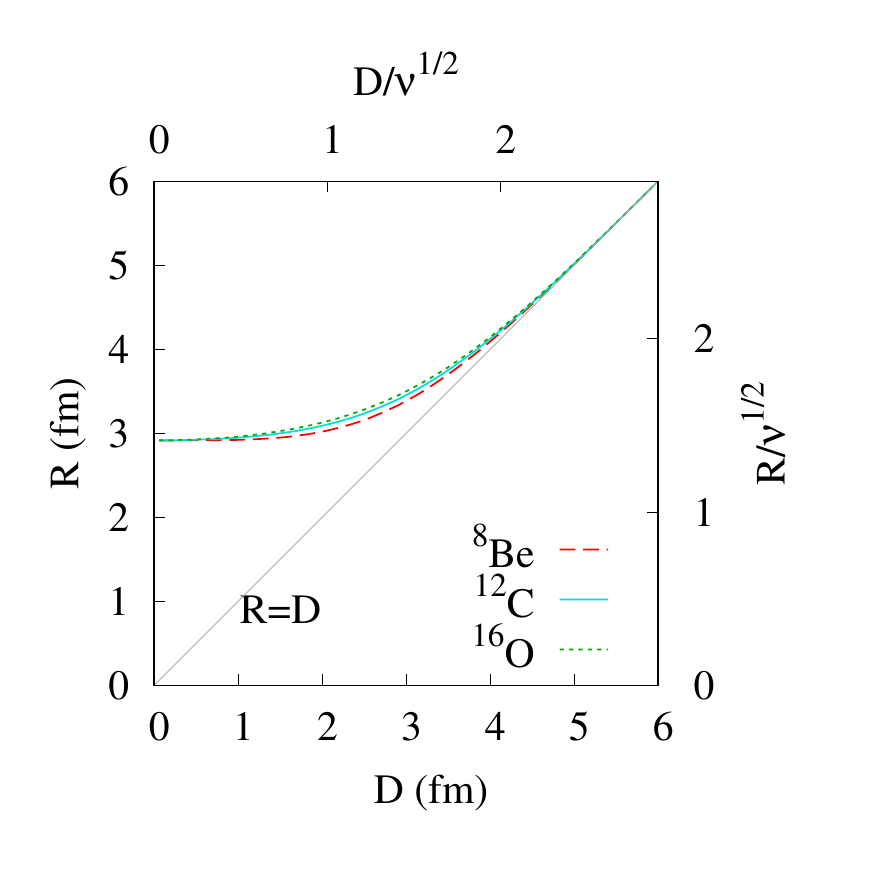}
\caption{Physical coordinates $R$ plotted
as functions of $D$ for the $n\alpha$ wave functions. 
%The units of the  bottom and left axes are
%fm, and those of the right and top axes are $\nu^{-1/2}$ 
%($\nu=0.235$ fm$^{-2}$).
\label{fig:dtow}}
\end{figure}

In the asymptotic region, where the antisymmetrization effect vanishes,
the parameter $D$ corresponds to
the mean distance between $\alpha$ positions. However,
this is not the case in the small $D$ region in which $D$ no longer has 
the physical meaning of the inter-cluster distance
because of the antisymmetrization effect between $\alpha$ clusters.
To demonstrate this antisymmetrization effect,
we follow the prescription for the transformation of coordinates 
proposed by Ono {\it et al.}~\cite{Ono:1992uy}.  Ono {\it et al.}
transformed a set of Gaussian centers $\{\bvec{Z}_i\}$ of single-nucleon wave functions
into a new set of coordinates $\{\bvec{W}_i\}$ in the  framework of antisymmetrized molecular dynamics~(AMD). 
They call the new coordinates  $\{\bvec{W}_i\}$  
``physical coordinates" and used them 
to avoid Pauli blocking in time-dependent AMD
to  study heavy-ion collision.
A detailed derivation of the new coordinates $\{\bvec{W}_i\}$
is provided in Appendix \ref{app:coordinates}.
In the case of the present $n\alpha$ model at a given value of $D$,
$\bvec{W}_i$ is analytically given as 
$\bvec{W}_i=\lambda_{n\alpha}(D)\bvec{Z}_i$ with the scaling factor
$\lambda_{n\alpha}(D)$. According to this transformation, the parameter $D$ is
transformed into a new coordinate $R=\lambda_{n\alpha}(D)D$, which can be regarded as
a ``physical coordinate" for the $\alpha$-$\alpha$ distance.
Figure \ref{fig:dtow} shows $R(D)$ for the $2\alpha$, $3\alpha$, and $4\alpha$ systems.
The $D$ dependence of $R$ shows almost no system dependence, indicating that
the antisymmetrization effect between two $\alpha$s is essential for $R(D)$.
In all cases, $R$ takes the minimum value of $\sqrt{2/\nu}$
in the $D\to 0$ limit, signifying that
two $\alpha$s can not come closer to each other
due to the Pauli blocking between identical nucleons in two $\alpha$s.
As $D$ increases, $R$ monotonically increases and approaches
$R\to D$ in the $D\gtrsim 5$~fm region.

%%%%%%%%%%%%%%%%%%%%%%%%%%%%%%
%%%%%%%%%%%%%%%%%%%%%%%%%

\subsubsection{Norm overlap and metric}

The norm kernel ${\cal N}(D,D')$ of the $2\alpha$ system
is given as
\begin{align}
{\cal N}(D,D')&=
\bratoket{\Phi_{2\alpha} (D)}{\Phi_{2\alpha} (D')}\nonumber\\
&=
\left[e^{-\nu\frac{(D'-D)^2}{4}}
-e^{-\nu\frac{(D'+D)^2}{4}}\right]^4.
\end{align}
In the asymptotic region of large $D$,
the $\alpha$-$\alpha$ relative motion is not affected by
antisymmetrization and is expressed by a Gaussian function of the relative coordinate $r$ as 
$e^{-\nu_D (r-D)^2}$, where $\nu_D =4\cdot 4\nu/(4+4)$.
Consequently, ${\cal N}(D,D')$ for $\epsilon\equiv D'-D$ becomes
a Gaussian function as
\begin{align}
{\cal N}(D,D+\epsilon)\to e^{-\frac{\nu_D}{2}\epsilon^2},
\end{align}
which satisfies the Gaussian overlap with constant width 
$\nu_D$.
The parameter $\nu_D$ is the metric
adopted in the GOA. It turns out 
that the norm kernel can be used as a measure to
evaluate the number of states contained
in the small interval $\epsilon$ of the parameter space and 
the metric is expressed by the leading  $\epsilon^2$ term of 
$1-{\cal N}(D,D+\epsilon)$ 
according to the $\epsilon^2$ expansion
\begin{align}
1-{\cal N}(D,D+\epsilon)\to \frac{\nu_D}{2}\epsilon^2+{\cal O}(\epsilon^4). 
\end{align}
We naively extend the prescription of this asymptotic feature of the norm kernel
and introduce 
a $D$-dependent metric $\gamma_N(D)$ 
as
\begin{align}
\gamma_{N}(D) \equiv \frac{1-{\cal N}(D,D+\epsilon)}
{(\nu_D/2) \epsilon^2}
\end{align}
where $\gamma_{N}(D)$ is normalized to 
$\nu_D$  to approach $\gamma_{N}(D)\to 1$ in the asymptotic region. 

%Here we use the notation
%to specify the metric derived from the $\gamma_{N}$
%norm overlap.

%%%%%%%%%%%%%%%%%%%%%%%%%%%%%%
\begin{figure}[!h]
\includegraphics[width=7.5 cm]{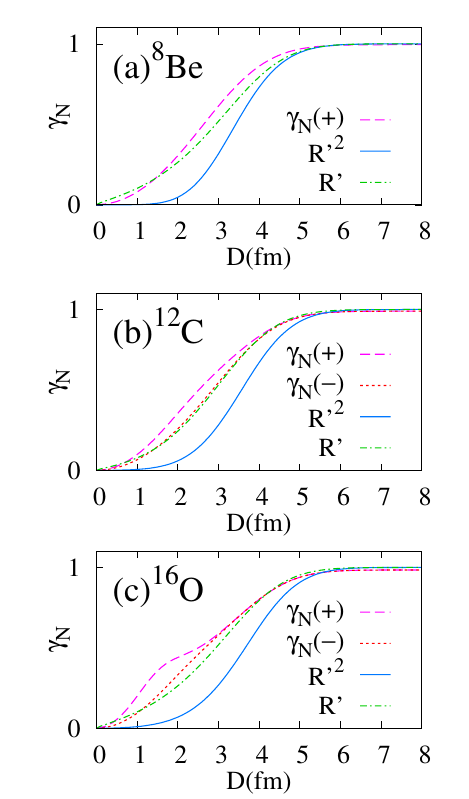}
\caption{$\gamma_N$ derived from the norm kernel ${\cal N}(D,D')$
for (a)$\Be8(+)$, (b)$\C12(\pm)$, and (c)$\O16(\pm)$. For comparison,
$R'=dR/dD$ and $(R')^2$ obtained from the physical coordinate $R$
are also shown.
\label{fig:gammaN}}
\end{figure}
%%%%%%%%%%%%%%%%%%%%%%%%%

In the present calculation, we consider the parity-projected
$n\alpha$ wave function at $D$ denoted as $\ket{D}=\ket{\Phi^\pm_{n\alpha}(D)}$
and redefine $\gamma_{N}(D)$ for the $n\alpha$ systems 
in the general form with the norm kernel
${\cal N}^\pm(D,D')=\bratoket{D}{D'}$ as
\begin{align}
\gamma_{N}(D) \equiv \frac{1-{\cal N}^\pm(D,D+\epsilon)}
{1-{\cal N}^\pm(D_\infty,D_\infty+\epsilon)},
\end{align}
where the
denominator is the asymptotic value in  the
$D\to \infty$ limit.
Figure \ref{fig:gammaN} shows the calculated values of $\gamma_N(D)$ 
for $\Be8(+)$, $\C12(\pm)$, and $\O16(\pm)$.
In the $D>5$ fm region, $\gamma_N\sim 1$, indicating that
the antisymmetrization effect almost vanishes in this region.

In the $D\lesssim 5$~fm region,
$\gamma_N$ becomes smaller than one as $D$ decreases
 and finally approaches zero at the $D\to 0$ limit because of the antisymmetrization effect. 
In $\C12$ and $\O16$, 
one can see a significant parity dependence of $\gamma_N$. 
In particular, $\gamma_N$ in $\O16(+)$ exhibits
unnatural oscillating behavior due to the parity-projection effect.
The intrinsic wave function of $4\alpha$ before parity projection is a mixed-parity state for $D\ne 0$, 
but it is a pure positive-parity state in the $D\to 0$ limit. 
In the small-$D$ region, the shift $D\to D+\epsilon$ involves a drastic change of the
parity mixing ratio, which has a non-trivial effect on $\gamma_N$ via 
${\cal N}^\pm(D,D')$ calculated using the parity-projected wave function. 
In other words, the non-trivial parity dependence of $\gamma_N$ originates from the 
quantum effect associated with the parity symmetry restoration. 

Let us discuss the behavior of $R'\equiv dR/dD$ of the 
physical coordinate $R$.
Provided that the metric is constant in the coordinate space of $R$, 
$(R')^2$ is naively expected to be an alternative metric.
In Fig.~\ref{fig:gammaN}, the values of $R'$ and $(R')^2$ are compared with $\gamma_N$.
$(R')^2$ is strongly suppressed in the $D\lesssim 5$ fm region, and inconsistent with $\gamma_N$.
However, $R'$ is in a better agreement with $\gamma_N$ 
but does not describe the parity dependence of $\gamma_N$ in $\C12$ and $\O16$.

\subsection{$N\alpha$ systems with complex parameter}

We extend the $n\alpha$ wave function by introducing the imaginary part of the coordinate $D$ as
$D \to D+iP/(2\hbar\nu)$. This extension is achieved by using the complex parameter for the Gaussian centers
instead of the real parameter $D$.
For instance, in the case of $2\alpha$,
the extended $n\alpha$ wave function is given by replacing the real parameter $\bvec{S}=(0,0,D)$ 
for the Gaussian centers 
$\bvec{S}_1=-\bvec{S}_2=\bvec{S}/2$ of two $\alpha$s as 
$\bvec{S}=(0,0,D+\frac{iP}{2\hbar\nu})$.

The real parameter $P$ introduced here corresponds to an imaginary
shift of $D$, and 
the $n\alpha$ state $\ket{D,P}$ can be written as
\begin{align}
\ket{D,P}=n_0(D,P) e^{i\frac{P}{2\hbar\nu} \frac{\partial}{\partial D}} \ket{D},
\end{align}
where $n_0$ is the normalization factor determined so that  $\bratoket{D,P}{D,P}=1$. 
Note that the operator  $e^{i\frac{P}{2\hbar\nu} \frac{\partial}{\partial D}}$ is not a unitary operator.
Let us consider the $2\alpha$ system. 
In the asymptotic region, where the antisymmetrization effect vanishes, 
the mean positions and momenta
of nucleons are given as $\braket{\bvec{r}_i}=(0,0,\pm D/2)$ and 
$\braket{\bvec{p}_i}=(0,0,\pm P/2)$, respectively, while those of
the relative motion between two $\alpha$s are given as
$\braket{\bvec{r}}=(0,0,D)$ and $\braket{\bvec{p}}=(0,0,\pm (\mu_{n\alpha}/M_N) P)$, respectively. 
Here $\mu_{n\alpha}$ is the reduced mass $\mu_{2\alpha}=M_N/2$ for the $2\alpha$ systems.
Indeed, the operator $e^{i\frac{P}{2\hbar\nu}\frac{\partial}{\partial D}}$ can be written using
the boost operators of nucleons with momenta $\pm P/2$ in the opposite
direction with an $D$- and $P$-dependent overall factor. 
In a similar way, the reduced mass for the $n\alpha$ systems is defined as 
$\mu_{n\alpha}= (D/d)^2 (M_N/A)$ using the nucleon mass $M_N$ and $d=|\bvec{S}_m|$.

We calculate the energy expectation value of the finite-momentum state $\ket{D,P}$ as 
\begin{align}
E_P={ \bra{D,P}\hat H\ket{D,P}},
\end{align}
and define the inverse mass $1/{M_P(D)}$ from the following relation,
\begin{align}\label{eq:massp}
\Delta E_P(D)=E_{P}(D)-E(D)=\frac{\hbar^2}{2M_P(D)}P^2,
\end{align}
where $E(D)=\bra{D}\hat H\ket{D}$ is the energy at $D$ and
$P=0$.

\begin{figure}[!h]
\includegraphics[width=6 cm]{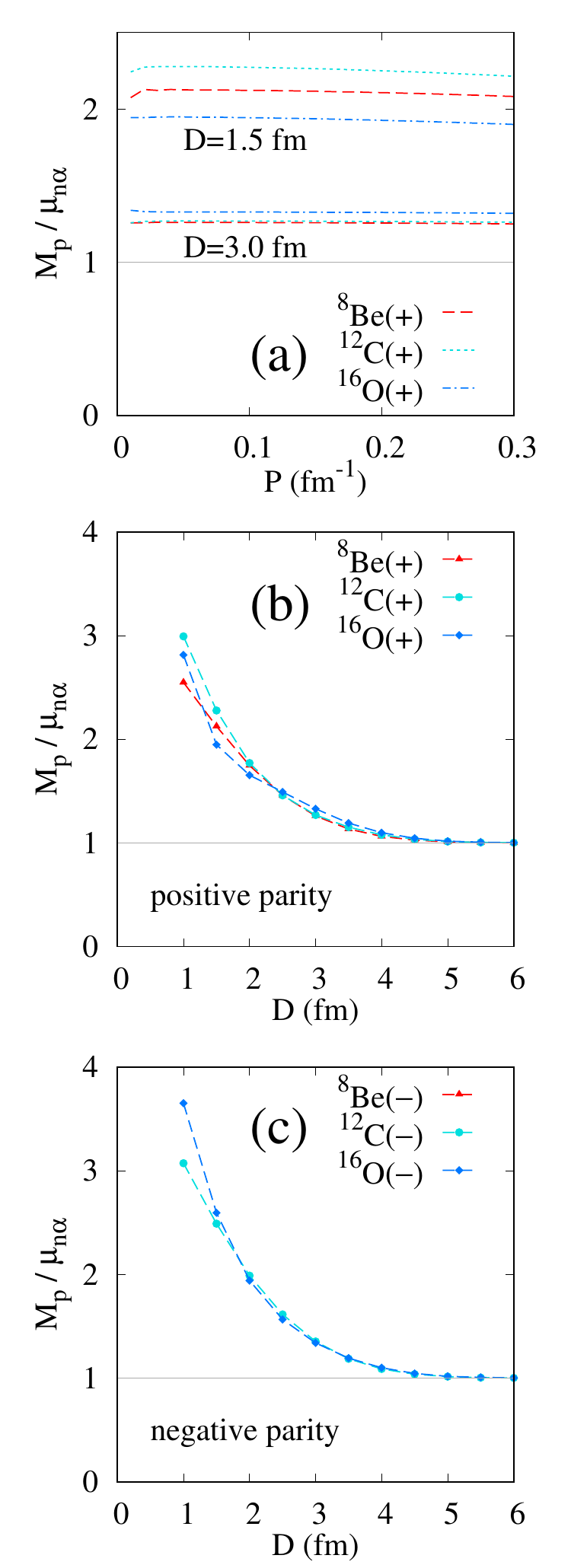}
\caption{Mass $M_P$ evaluated by $\ket{D,P}$ with the complex parameter $D+iP/(2\hbar\nu)$.
(a) $P$ dependence of $M_P$ at $D=1.5$~fm and 3.0~fm,
(b) $D$ dependence of $M_P$ for positive-parity states, and (c) $D$ dependence of Mp for negative-parity states.
The values are divided by the asymptotic value $\mu_{n\alpha}=(D/d)^2 (M_N/A)$
of $n\alpha$ systems.
\label{fig:mass}}
\end{figure}

Figure \ref{fig:mass} shows the $P$ and $D$ dependences of $M_P$.
As shown in Fig.~\ref{fig:mass}(a) for the $P$ dependence,  
$M_P(D)$ is almost constant in the $P\le 0.3$~fm$^{-1}$ region, and therefore we omit
the $P$ dependence of $M_P(D)$ in the following discussion.
The $D$ dependence of $M_P$ at  $P=0.05$~fm$^{-1}$ for the positive- and negative-parity states
is presented in Figs.~\ref{fig:mass}(b) and (c), respectively. 
The values relative to the asymptotic value
$\mu_{n\alpha}$ are plotted.
In the $D>5$~fm region, $M_P/\mu_{n\alpha}$ is approximately
equal to 1, indicating that the antisymmetrization effect almost vanishes
in this region.
As $D$ decreases, $M_P$ increases  in the $D\lesssim 5$~fm region.
This increasing behavior of $M_P$ seems inconsistent with the naive expectation that
the antisymmetrization, {\it i.e.}, the Pauli blocking effect, may give a repulsive effect and
contributes to reducing the inertial mass of the kinetic term.
As described above, 
$M_P$ is measured by the inverse of the energy difference $\Delta E_P$
between two states $\ket{D}$ and $\ket{D,P}$. Since 
the antisymmetrization effect suppresses the state difference  in the interior region, 
it contributes to decreasing the energy difference and increasing $M_P$.
Therefore, it may not be adequate to directly use the obtained $M_P$ values 
as the inertial mass of the collective model, but 
some modification may be necessary by taking into account the
antisymmetrization effect.

%%%%%%%%%%%%%%%%%%%%%%%%%%%%%%
%%%%%%%%%%%%%%%%%%%%%%%%%

\subsection{Small-amplitude description}

In the case of $P\ne 0$,
the state $\ket{D,P}$ contains
component orthogonal to $\ket{D}$. 
However, in the $n\alpha$ model, 
the two-dimensional GCM using $(D,P)$ obtains 
results consistent with the results of one-dimensional GCM with $D$
for low-lying bound states,
because the model space of $(D,P)$ contains redundant states.
%This redundancy is a general feature of Gaussian wave packet frameworks
%as the imaginary shift of the Gaussian center can be practically expressed 
%by the real shift of the Gaussian center. 
Nevertheless, one of the advantages of introducing the complex parameter
is that $\ket{D,P}$
provides a semi-classical picture of oscillation around the coordinate $D$
in a simple expression of the single-basis wave function at $(D,P)$.
In particular, the ${\cal O}(P)$ term of $\ket{D,P}$ involves
the time-odd components for the small-amplitude mode
around the static solution $\ket{D}$.

\begin{figure}[!h]
\includegraphics[width=8.5 cm]{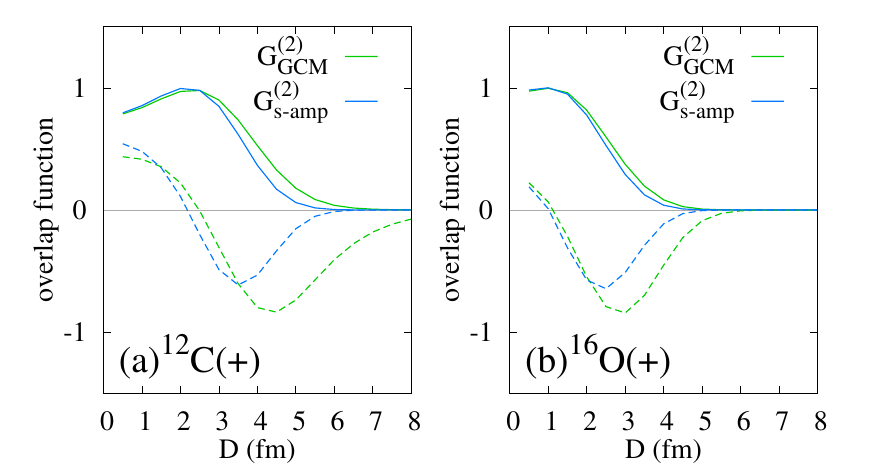}
\caption{Overlap function $G^{(2),\pm}_{\textrm{s-amp},k}$ obtained by two-basis diagonalization
for the small-amplitude calculation of (a)$\C12(+)$ and (b)$\O16(+)$.
The GCM results of the overlap function, $G^{(2),\pm}_{\textrm{GCM},k}$, are also shown
for comparison. Solid (dashed) lines indicate the overlap functions of lower $(\pm)_1$ [higher $(\pm)_2$] states.
\label{fig:dia2}}
\end{figure}

To discuss the vibration feature of the radial excitation
in the $\C12$ and $\O16$ systems,
we take a small value of $P=0.1$ fm$^{-1}$ at the optimized
$D_0$ for the energy minimum of $E(D)$, 
and diagonalize two bases of the time reversal partners, $\ket{D_0,P}$ and $\ket{D_0,-P}$,
to obtain a small-amplitude oscillation in the ground and excited states, $\Psi^\pm_{\textrm{s-amp},k}$.
The results obtained by the two-basis diagonalization for the small-amplitude approximation
are shown in the third column of Table \ref{tab:rmsr} for the energy and radii, and in Fig.~\ref{fig:dia2}
for the overlap function $G^{(2),\pm}_{\textrm{s-amp},k}(D)\equiv \bratoket
{\Phi^\pm_{n\alpha} (D)}{\Psi^\pm_{\textrm{s-amp},k}}$.
Compared with the
GCM calculation, the small-amplitude calculation tends to overestimate the energies and underestimate the radii,   
indicating that these states obtained by the GCM are not 
small-amplitude vibrations but large-amplitude motion. 
In particular, significant differences from the GCM results
are obtained for the excited states shown in Fig.~\ref{fig:dia2}.  
An exception is the $\O16(+)_1$ state, which is
well reproduced by the small-amplitude approximation.

%%%%%%%%%%%%%%%%%%%%%%%%%%%%%%
%%%%%%%%%%%%%%%%%%%%%%%%%

\subsection{Collective Hamiltonian}

In general, it is difficult for non-microscopic potential models 
to obtain results
equivalent to microscopic calculations.
Nevertheless, semi-microscopic or phenomenological potential models
are useful to
obtain reasonable results and are widely applied to study the dynamics of cluster motion.
The aim of this section is to construct a collective Hamiltonian 
 that can approximately
describe the fundamental properties of the cluster motion in
the ground and excited states obtained by the microscopic calculation of the GCM.
At small $D$, the microscopic state $\ket{D}$ is a highly non-localized state 
and contains strong
quantum effects such as  antisymmetrization and parity-projection, 
and therefore the GOA is not applicable.
We consider an alternative approach as follows.

The basic idea is that we assume local collective variables in the collective Hamiltonian
by utilizing diagonal elements, {\it i.e.},
expectation values of microscopic operators ${\hat O}$ obtained by
a single basis of the microscopic $n\alpha$ model wave function. 
This signifies that
non-trivial microscopic effects, such as the antisymmetrization and parity projection, 
are taken into account as local inputs as much as possible.
In the asymptotic region, $D$, $M_P(D)$, and $g^\pm_{\textrm{GCM},k}(D)$ can be regarded as
the collective coordinate, mass, and collective wave function of the radial motion of the
$n\alpha$ systems.
Indeed, $D$ and $M_P(D)$ satisfy the asymptotic conditions, 
$\braket{\bvec{r}}\to D$ and $M_P\to \mu_{n\alpha}$, respectively. 
Moreover, $|g^\pm_{\textrm{GCM}_k}(D)|$ represents the probability, and $g^\pm_{\textrm{GCM}_k}(D)$
satisfies the orthonormal condition in the 
coordinate $D$ space as given in Eq.~\eqref{eq:g-orthonormal}.
We start from the collective Hamiltonian with
the collective coordinate $D$ space and the mass $M_P$ by taking into account 
the metric in the $D$ space, and consider several options. 
Then we solve the eigenvalue problem of the collective model and evaluate
whether it provides results in reasonable agreement with the microscopic results of the GCM.

We assume that 
the collective Hamiltonian consists of the kinetic and potential terms as
\begin{align}
{\cal H}_\textrm{coll}={\cal T}_\textrm{coll}+{\cal V}_\textrm{coll}(D),
\end{align}
and suppose that it can approximately describe energies $E^\pm_{\textrm{GCM},k}$ and collective wave functions $g^\pm_{\textrm{GCM},k}(D)$.
%\subsubsection{Potential term}
For the potential term, we adopt a parity-dependent local potential of
${\cal V}_\textrm{coll} (D)=E^\pm (D)-T_0$, where $T_0$ is a constant value of $T_0=\hbar\omega/4$
$(\omega=2\hbar^2\nu/M_N)$ 
for the zero-point energy contained in the microscopic $n\alpha$ wave function at a large $D$.

In general, the coordinate $D$ space has a metric $\gamma(D)$, and
observables for local operators ${\cal O}_\textrm{coll}$ in the $D$ space are given by the
expectation values using the collective wave function $\Phi_\textrm{coll}(D)$
and the weight factor $\sqrt{\gamma}$ as
\begin{align}\label{eq:O-coll}
\braket{ {\cal O}_\textrm{coll} }=\int \Phi_\textrm{coll}^*(D){\cal O}_\textrm{coll}(D) \Phi_\textrm{coll}(D) \sqrt{\gamma}dD.
\end{align}
Following a prescription for quantization in one dimension with the metric $\gamma$,
we introduce the kinetic term of ${\cal H}_\textrm{coll}$ as
\begin{align}\label{eq:coll-kin}
{\cal T}_\textrm{coll}=-\frac{\hbar^2}{2} \frac{1}{\sqrt{\gamma}} \frac{d}{d D} \sqrt{\gamma}\frac{1}{\gamma\mu}
\frac{d}{d D},
\end{align}
%$\gamma$ is normalized to $\nu_D$, and therefore, so that the kinetic term
The microscopic effects are taken into account in the $D$ dependence of $\gamma$ and $\mu$, 
but in the asymptotic region, 
they should be constant as 
$\gamma\to 1$ and $\mu\to \mu_{n\alpha}$ so that 
the kinetic term takes
the standard form;
\begin{align}
{\cal T}_\textrm{coll}\to -\frac{\hbar^2}{2\mu_{n\alpha}}\frac{d^2}{dD^2} \quad (D\to \infty).
\end{align}

For the metric $\gamma$ and mass $\mu$ in the collective model, we consider five cases as follows.
In the first case (1), we adopt the $D$-dependent metric and mass as
$\gamma=\gamma_{N}(D)$ and $\mu=M_P(D)$, which are obtained by utilizing
the norm kernel ${\cal N}(D,D')$ and the finite-momentum state $\ket{D,P}$.
Note that $\gamma_{N}(D)$ and $M_P(D)$ are parity dependent as they are obtained with
the parity-projected $n\alpha$ wave function as mentioned previously.
In the second (2), $\gamma=1$ is kept to be constant, and
we use the mass $\mu=\gamma_{N}(D)M_P(D)$.
%This corresponds to microscopic effects are effectively incorporated only in the inertial mass.
In the third case (3),
we use the mass $\mu=M_P(D)$ and take an alternative metric
derived from the physical coordinate $R(D)$ as
$\gamma=dR/dD\equiv \gamma_R$. In this case, 
$\mu$ is parity-dependent, but $\gamma$ is not. 
The fourth case (4) is a reference case;
we use the naive ansatz of the constant values $\gamma=1$ and $\mu=\mu_{n\alpha}$.
We also perform a test calculation in the fifth case (5) using $\gamma=1$ and $\mu=M_P(D)$.
It should be noted that all cases satisfy the asymptotic conditions.

In Table~\ref{tab:gamma}, we summarize 
the five sets of $\gamma$ and $\mu$ in the collective model, 
which are labeled cal(1), (2), (3), (4), and (5).
In the table, the notations $\gamma^\pm_N$ and $M^\pm_P$ are used to explicitly
denote the parity dependence of $\gamma_N$ and $M_P$.
It should be noted that the $D$-dependent $\gamma$ incorporates
the microscopic effects on the coordinate $D$ space from the microscopic wave function,
but not the dynamical effect from the microscopic Hamiltonian. However
$M_P$ in the kinetic term 
and the potential term ${\cal V}_\textrm{coll}$ incorporate the
dynamical effects in addition to the microscopic effects from the microscopic wave function.

%%%%%%%%%%%%%%%%%%%%%%%%%%%%%%
\begin{table}[!ht]
\caption{Values of $\gamma$ and $\mu$ used in the kinetic term
Eq.~\eqref{eq:coll-kin} of the collective Hamiltonian ${\cal H}_\textrm{coll}$.
The boundary condition of $\Phi^{-}_\textrm{coll}$ at $D=0$ for negative-parity states is
also listed; the default condition $\Phi^{-\prime}_\textrm{coll}(0)=0$ in cal(1,2,3,4,5) and the optional condition $\Phi^-_\textrm{coll}(0)=0$ 
in cal(1b,2b) cases, which are noted as $\Phi'$ and $\Phi$, respectively.
\label{tab:gamma}
}
\begin{ruledtabular}
\begin{tabular}{lccccccc}
default sets & cal(1) & cal(2) & cal(3) & cal(4) & cal(5) \\
\hline
$\mu$ & $M^\pm_P$ & $\gamma^\pm_NM^\pm_P$ & $M^\pm_P$ & $\mu_{n\alpha}$ & $M^\pm_P$ \\
$\gamma$ & $\gamma^\pm_N$ & $1$ & $\gamma_R$ & 1 & 1 \\
$\Phi^-_\textrm{coll}(D=0)$& $\Phi'$& $\Phi'$& $\Phi'$& $\Phi'$& $\Phi'$ \\
optional sets& cal(1b) & cal(2b) & & & \\
$\Phi^-_\textrm{coll}(D=0)$& $\Phi$& $\Phi$& & & \\
\end{tabular}
\end{ruledtabular}
\end{table}
%%%%%%%%%%%%%%%%%%%%%%%%%%%%

\subsection{Collective wave function}
The collective wave function
$\Phi_\textrm{coll}(D)$ is obtained by solving the eigenvalue problem of the collective
Hamiltonian ${\cal H}_\textrm{coll}$ in the coordinate space $D$
under the orthonormal condition
\begin{align}
&\bratoket{\Phi_\textrm{coll}(D)} {\Phi_\textrm{coll}(D)}\nonumber\\
&=\int \Phi_\textrm{coll}^*(D)\Phi_\textrm{coll}(D) \sqrt{\gamma} dD=1.
\end{align}
The obtained eigenvalue of the collective Hamiltonian is the energy of the collective state
as follows
\begin{align}
&\bra{\Phi_\textrm{coll}(D)} {\cal H}_\textrm{coll} \ket {\Phi_\textrm{coll}(D)}\nonumber\\
&=\int \Phi_\textrm{coll}^*(D){\cal H}_\textrm{coll} \Phi_\textrm{coll}(D) \sqrt{\gamma}dD.
\end{align}
The root-mean-square radii are calculated with Eq.~\eqref{eq:O-coll}
by assuming that the collective operator
${\cal O}_\textrm{coll}(D)$ is given by the diagonal element (expectation value)
of the microscopic wave function at $D$ as
\begin{align}
{\cal O}_\textrm{coll}(D)=
\bra{\Phi^\pm_{n\alpha} (D)} \sum_i (\hat{\bvec{r}}_i-\hat{\bvec{r}}_G)^2 \ket{\Phi^\pm_{n\alpha} (D)},
\end{align}
where $\bvec{r}_G$ is the total center of mass coordinate.
We define $\phi_\textrm{coll}(D)\equiv \gamma^{1/4} \Phi_\textrm{coll}(D)$,
which satisfies $\int \phi^*_\textrm{coll}(D) \phi_\textrm{coll}(D) dD=1$
to compare the collective wave functions with the GCM solution
$g^\pm_\textrm{GCM}(D)$.
% $g^\pm_\textrm{GCM}(D)\equiv G_{\textrm{GCM}}(D)/\sqrt{\Delta_D} $

%\subsubsection{Boundary condition}

The boundary condition of the collective wave function at $D=0$ is not trivial
because of the antisymmetrization effect. 
For positive-parity states, we set $d\Phi^+_\textrm{coll}(D)/dD=\Phi^{+\prime}_\textrm{coll}(D)=0$
at $D=0$.
For negative-parity states, the GCM amplitude function $g^-_\textrm{GCM} (D)$
is inconsistent with the standard condition $\Phi^-_\textrm{coll}(D)=0$ of negative-parity states, 
that is, the parity transformation does not correspond to the transformation 
$g^\pm_\textrm{GCM} (D)\to g^\pm_\textrm{GCM} (-D)$ in the collective coordinate $D$ space. 
Instead,  we choose the condition $\Phi^{-\prime}_\textrm{coll}(D)=0$ at $D=0$, with which
we can obtain a better result than with the standard choice, as shown later.
This condition corresponds to a calculation with the same condition as the positive-parity states
but with the parity-dependent Hamiltonian. In other words, the parity-projection effects are incorporated
in the Hamiltonian but not in the $D=0$ boundary condition. 
In the asymptotic region, we adopt the same
bound-state approximation used in the GCM calculation. Namely,
the collective wave function is expressed by a sum of localized
Gaussians with center positions from 0.5~fm to 8.0~fm with intervals of $0.5$~fm,
and the eigenvalue problem is solved by diagonalization.

\subsection{Results of the collective Hamiltonian model}

%%%%%%%%%%%%%%%%%%%%%%%%%%%%%%
\begin{figure}[!htpb]
\includegraphics[width=8 cm]{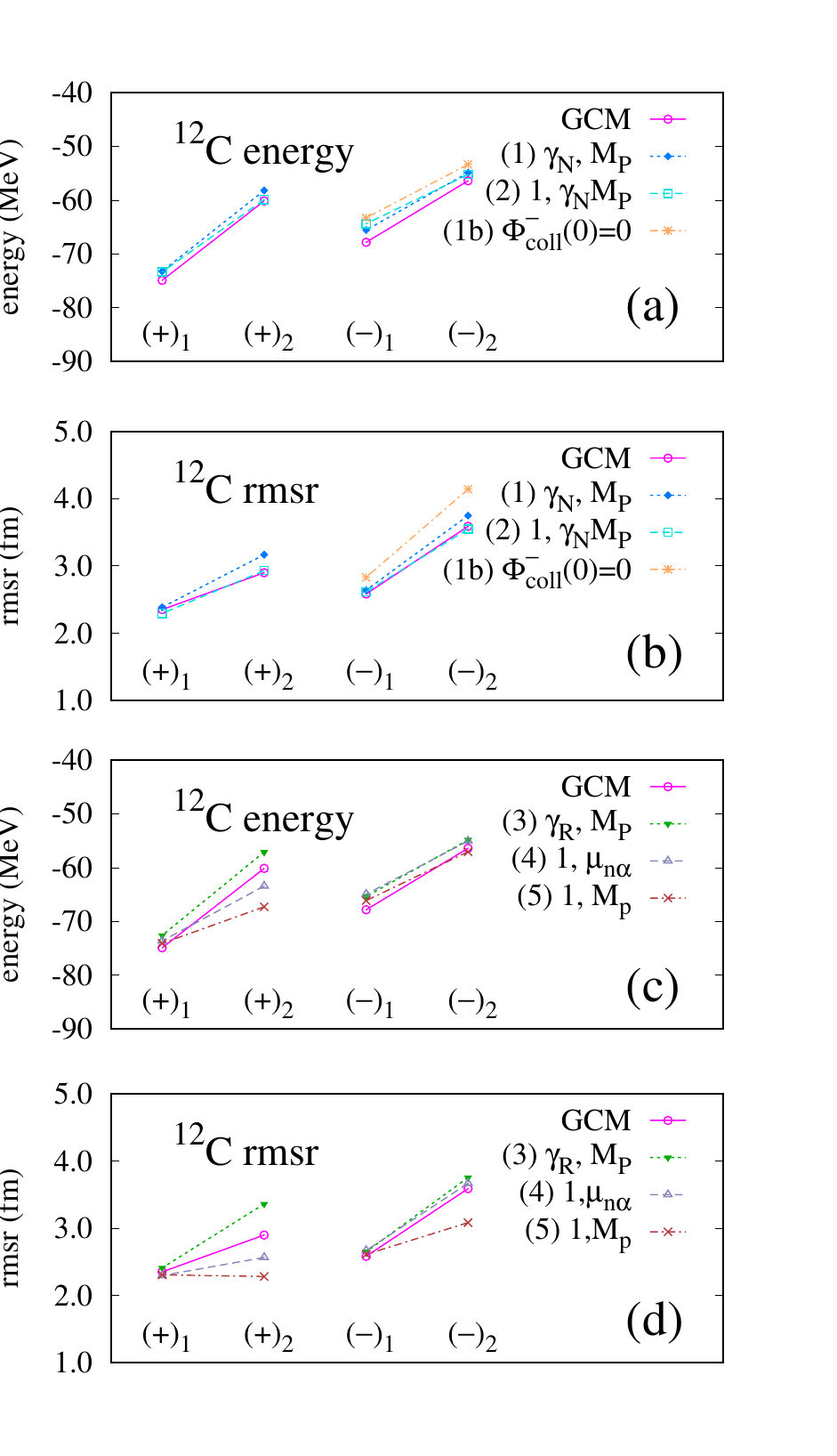}
\caption{Energies and radii~(rmsr) of $\C12$
calculated by the
collective models in comparison with the GCM results.
This figure presents the results of cal(1), cal(2), and cal(1b) 
for the (a) energies and (b) radii, and the results of cal(3), cal(4), and cal(5) for
the  (c) energies and (d) radii.
\label{fig:spe-rmsr-c12}}
\end{figure}
%%%%%%%%%%%%%%%%%%%%%%%%%

%%%%%%%%%%%%%%%%%%%%%%%%%%%%%%
\begin{figure}[!htpb]
\includegraphics[width=8 cm]{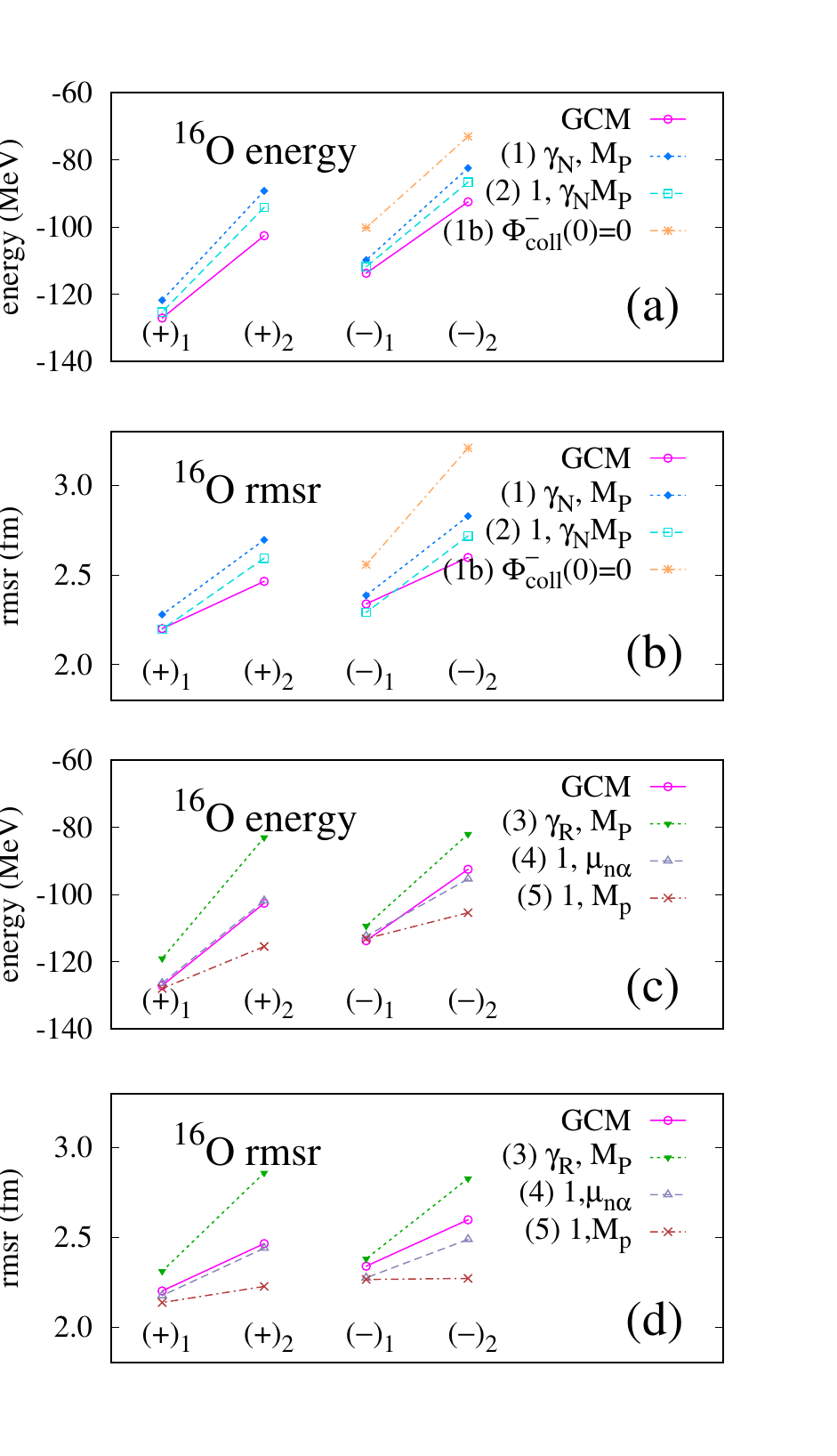}
\caption{Energies and radii~(rmsr) of $\O16$
calculated by the
collective models in comparison with the GCM results.
This figure presents the results of cal(1), cal(2), and cal(1b) 
for the (a) energies and (b) radii, and the results of cal(3), cal(4), and cal(5) for
the  (c) energies and (d) radii.
\label{fig:spe-rmsr-o16}}
\end{figure}
%%%%%%%%%%%%%%%%%%%%%%%%%

%%%%%%%%%%%%%%%%%%%%%%%%%%%%%%
\begin{figure}[!htpb]
\includegraphics[width=7 cm]{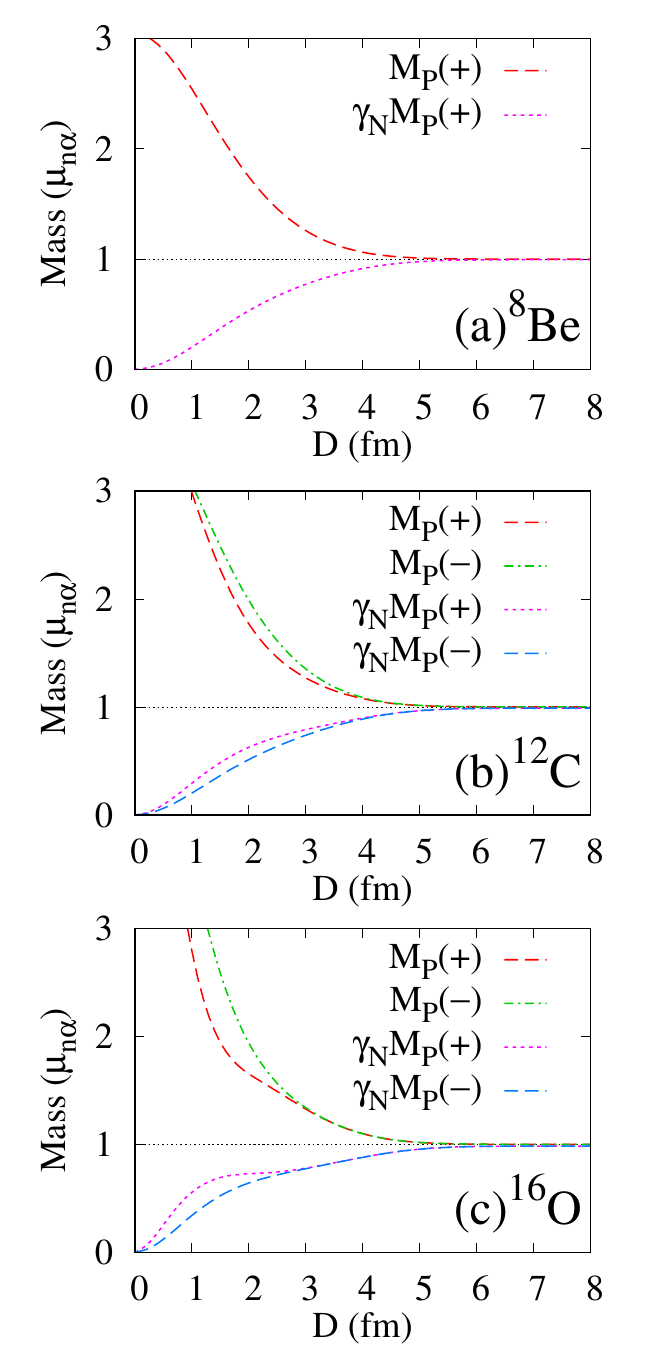}
\caption{$D$ dependences of the inertial mass $\mu=M_P$
and $\gamma_N M_P$ used in cal(5) and (2) for (a)$\Be8(+)$,
(b)$\C12(\pm)$, and (c)$\O16(\pm)$. 
The values relative to the
constant mass $\mu=\mu_{n\alpha}$ corresponding to cal(4) are plotted.
\label{fig:massg}}
\end{figure}
%%%%%%%%%%%%%%%%%%%%%%%%%

We show the results of the lowest and first excited states
obtained by the collective model of the five cases 
and compare them with the microscopic results of the GCM.
The results for the energies and radii of the $(+)_{1,2}$ and $(-)_{1,2}$ states of
$\C12$ are shown in Fig.~\ref{fig:spe-rmsr-c12}, while those of $\O16$ are shown in
Fig.~\ref{fig:spe-rmsr-o16}.
As shown in Figs.~\ref{fig:spe-rmsr-c12}(a), \ref{fig:spe-rmsr-c12}(b), \ref{fig:spe-rmsr-o16}(a), and \ref{fig:spe-rmsr-o16}(b),
the collective model cal(1) using $\gamma=\gamma_N$ and $\mu=M_P$
reasonably reproduces the GCM result of the energies and radii of the
lowest and excited states $\C12(\pm)_{1,2}$ and $\O16(\pm)_{1,2}$.
The second model cal(2) using $\gamma=1$ and $\mu=\gamma_NM_P$
produces similar results to the cal(1) results for $\C12$ and somewhat better results  for
$\O16$.

Let us compare the results
obtained by the optional case of the negative-parity boundary condition 
cal(1b) for $\Phi^-_\textrm{coll}(0)=0$ with the cal(1) results for $\Phi^{-\prime}_\textrm{coll}(0)=0$. 
The former calculation (1b) overestimates the energies and 
radii of the GCM results, indicating that
the condition $\Phi^-_\textrm{coll}(0)=0$ is not appropriate for the collective wave functions in the 
$D$ space.

Other model calculations of cal(3), (4), and (5) are not satisfactory
in systematically reproducing the GCM results~[see Figs.~\ref{fig:spe-rmsr-c12}(c), \ref{fig:spe-rmsr-c12}(d), \ref{fig:spe-rmsr-o16}(c), and \ref{fig:spe-rmsr-o16}(d)].
In particular, these calculations failed to reproduce the properties of the $\C12(+)_2$ state, 
and while the calculations of cal(3) and (5) cannot describe the  $\O16(+)_2$ state.
The model cal(3) tends to overestimate the radii of the $\C12(+)_2$ and $\O16(+)_2$
states, because the metric $\gamma_R$ used in cal(3) is slightly smaller than $\gamma_N$
for the positive-parity states 
and provides a stronger repulsive effect in the kinetic term
than that in the case of cal(1).
Compared with cal(3) and (5), 
 improved results are obtained by cal(4) for some states.
However, the results of cal(4) are not global reproduction but the agreement is  
state- and system-dependent. 
Therefore,
the sets of $\gamma$ and $\mu$ used in these models 
do not work for describing the collective motion along $D$ in the $n\alpha$ systems.

We compare the results obtained by cal(4) and cal(5) with those of cal(2).
These three calculations use the constant metric $\gamma=1$ but different values of 
the collective mass in the kinetic term; $\mu=\gamma_N M_P$, $\mu_{n\alpha}$, and $M_P$ 
are used in cal(2), cal(4), and cal(5), respectively.
Figure \ref{fig:massg} shows the $D$ dependence of $\mu$ of cal(2) and cal(5) 
relative to the constant mass $\mu_{n\alpha}$ for cal(4).
In the interior region, $\mu=\gamma_N M_P $ for cal(2) is suppressed
because of the antisymmetrization effect, whereas $\mu=M_P$ for cal(5) is enhanced.

Here, the cal(4) underestimates the radius of the
$\C12(+)_2$ state because the constant $\gamma$ and $\mu$ values of cal(4) provide
no repulsive effect in the kinetic term compared with the case of cal(2).
The model cal(5) significantly underestimates the radii of
all states of $\O16$ as well as the $\C12(\pm)_2$ states because
$\mu=M_P$, which is largely enhanced in the interior region, provides
more attractive effects compared with the cases cal(2) and cal(4).

%%%%%%%%%%%%%%%%%%%%%%%%%%%%%%
\begin{figure}[!htpb]
\includegraphics[width=8.5 cm]{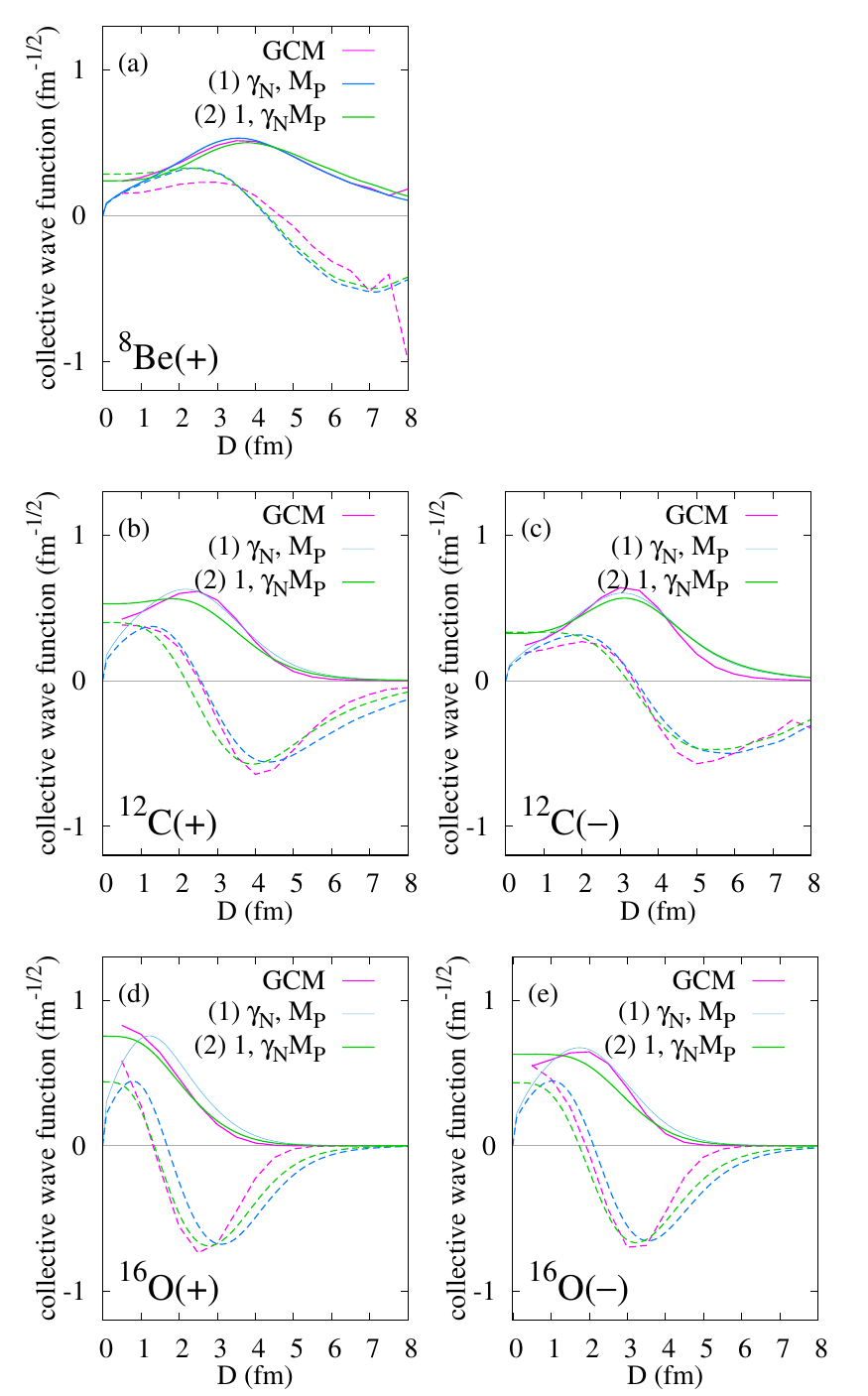}
\caption{Collective wave functions $\phi_\textrm{coll}(D)$ calculated
by the collective models of cal(1) and cal(2)
compared with the GCM amplitude
function $g^\pm_\textrm{GCM}(D)$.
The results are displayed for
(a)$\Be8(+)$, (b)$\C12(+)$, (c)$\C12(-)$, (d)$\O16(+)$, and (e)$\O16(-)$.
Solid (dashed) lines indicate the functions of lower $(\pm)_1$ [higher $(\pm)_2$] states.
\label{fig:dia1-coll1}}
\end{figure}
%%%%%%%%%%%%%%%%%%%%%%%%%

%%%%%%%%%%%%%%%%%%%%%%%%%%%%%%
\begin{figure}[!htpb]
\includegraphics[width=8.5 cm]{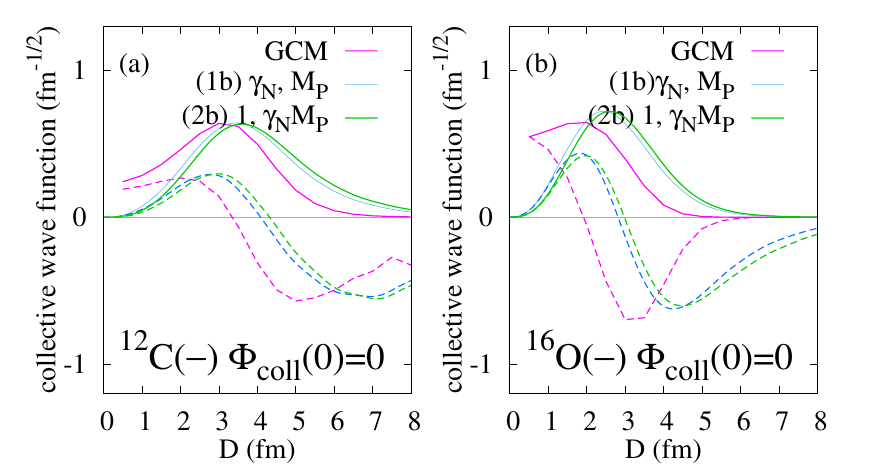}
\caption{Collective wave functions $\phi_\textrm{coll}(D)$ calculated
by the collective models cal(1b) and cal(2b) for the optional case of the negative-parity 
boundary condition
$\Phi^-_\textrm{coll}(0)=0$, in comparison 
with the GCM amplitude function $g^\pm_\textrm{GCM}(D)$.
The results for (a)$\C12(-)$ and (b)$\O16(-)$.
Solid (dashed) lines indicate the functions of lower $(-)_1$ [higher $(-)_2$] states.
\label{fig:dia1-np-coll1}}
\end{figure}
%%%%%%%%%%%%%%%%%%%%%%%%%

%%%%%%%%%%%%%%%%%%%%%%%%%%%%%%
\begin{figure}[!htpb]
\includegraphics[width=8.5 cm]{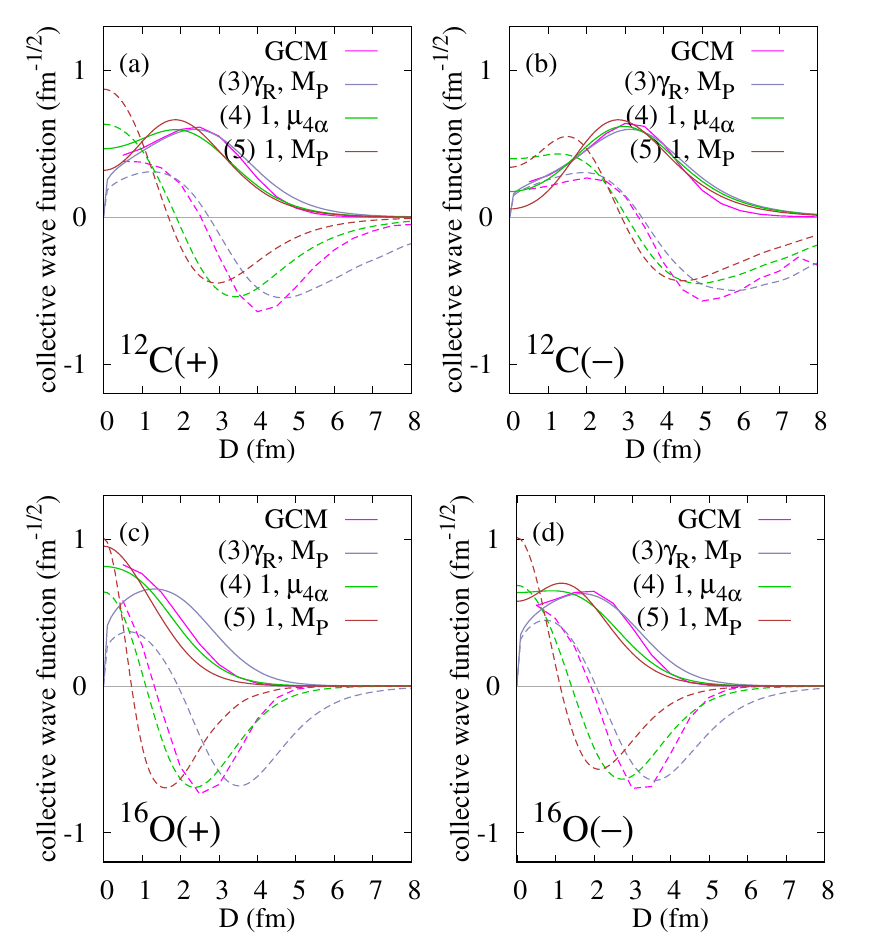}
\caption{
Collective wave functions $\phi_\textrm{coll}(D)$ calculated
by the collective models  cal(3), cal(4), and cal(5) for
(a) $\C12(+)$, (b) $\C12(-)$, (c) $\O16(+)$, and
(d) $\O16(-)$, in comparison with the GCM amplitude
function $g^\pm_\textrm{GCM}(D)$.
Solid (dashed) lines indicate the functions of lower $(\pm)_1$ [higher $(\pm)_2$] states.
\label{fig:dia1-coll2}}
\end{figure}
%%%%%%%%%%%%%%%%%%%%%%%%%

To examine the behavior of the collective motion in greater detail,  Fig.~\ref{fig:dia1-coll1} exhibits
the collective wave functions $\phi_\textrm{coll}$ obtained by the collective models of 
cal(1) and cal(2) compared with the GCM amplitude function $g^\pm_\textrm{GCM}$.
Although
the same boundary condition at $D=0$ is adopted for $\Phi_\textrm{coll}(D)$ in cal(1) and cal(2),
$\phi_\textrm{coll}(D)=\gamma^{1/4}\Phi_\textrm{coll}(D)$
has different behavior at $D=0$. In the case of cal(1),  an additional node appears in $\phi_\textrm{coll}$ at $D=0$, because $\gamma_N(D)\to 0$
in the $D\to 0$ limit, but 
not in the case of cal(2). 
Because of this additional node, $\phi_\textrm{coll}$ of cal(1) slightly shifts 
outward compared with the cal(2) result. In particular, in the $\O16(+)$ states,  
the result of cal(1) fails to describe the concentration around $D\sim 0$ of 
the GCM amplitude
in the deep potential, because the $D=0$ node prevents
$\phi_\textrm{coll}$ from penetrating in the short distance region
[see Fig.~\ref{fig:dia1-coll1}(d)].

In principle, $\phi_\textrm{coll}(D)$ of cal(1) and cal(2) are similar to each other except for the 
$D=0$ node in cal(1), which 
%suggesting that microscopic effects on the metric can be approximately
%renormalized in the effective mass as $\mu=\gamma_N M_P $ as done in cal(2).
does not satisfy our requirement of reproducing the GCM amplitude function.
The set of $\gamma=1$ and $\mu=\gamma_N M_P $ in cal(2) is a simple prescription
to effectively take into account the antisymmetrization effect in the collective mass of the kinetic term
by avoiding this unfavorable condition of the $D=0$ node.

The collective wave functions of cal(1b) and (2b) for the optional choice of the negative-parity condition 
are shown in Fig.~\ref{fig:dia1-np-coll1}. The condition $\Phi^-_\textrm{coll}(0)=0$
strongly suppresses $\phi_\textrm{coll}(D)$ in the interior region and is not suitable for reproducing the 
GCM amplitude functions.

The collective wave functions of other calculations, cal(3), (4), and (5),
are shown in Fig.~\ref{fig:dia1-coll2}.
The differences in $\phi_\textrm{coll}$ of cal(3), (4), and (5) from the cal(1) and cal(2) results can be easily 
understood by the differences in the $D$ dependences of $\gamma$ and $\mu$ in the interior region. 
For example, 
the repulsive effect of antisymmetrization is too strong in cal(3) because $\gamma_R$ is smaller 
than $\gamma_N$, as shown in Fig.~\ref{fig:gammaN}, whereas it is too weak in cal(5) 
as expected from the enhanced $\mu=M_P$ as shown in Fig.~\ref{fig:massg}. 

From those analyses of the collective model calculations,
it is concluded that the set $\gamma=1$ and $\mu=\gamma_N M_P$ of cal(2)
seems to be the best and simple choice among the five choices of the collective model
for the global reproduction of the cluster motion obtained by the GCM. 
This collective model corresponds to
a prescription for the derivation of the collective Hamiltonian
from the energy expectation value measured by the $\ket{D,P}$ state
\begin{align}
E_{P}(D)=P\frac{\hbar^2}{2M^\pm_P(D)}P+E^\pm(D)
\end{align}
as
\begin{align}\label{eq:coll-cal2}
&{\cal H}_\textrm{coll}\nonumber\\
&=\left(i\frac{d}{d D}\right) \frac{1}{\gamma^\pm_N}
\frac{\hbar^2} {2 M^\pm_P (D)} \left(i\frac{d}{d D}\right) -T_0+ E^\pm(D).
\end{align}
%\begin{align}\label{eq:coll-cal2}
%{\cal H}_\textrm{coll}=-\frac{d}{d D}\frac{1}{\gamma^\pm_N}
%\frac{\hbar^2} {2 M^\pm_P (D)} \frac{d}{d D}-T_0+ E^\pm(D).
%\end{align}
and the matrix element of a collective operator ${\cal O}_\textrm{coll}$ as
\begin{align}
\braket{ {\cal O}_\textrm{coll} }=\int \Phi_\textrm{coll}^*(D){\cal O}_\textrm{coll}(D) \Phi_\textrm{coll}(D) dD.
\end{align}
This model can properly describe the collective motion of $n\alpha$ systems
and approximately
reproduce the GCM results for the energies, radii, and amplitude functions. 
$\mu={\gamma_N}M_P$ in the kinetic term is regarded as the effective collective mass,
in which microscopic effects
such as antisymmetrization and parity projection on the model space
$\ket{D}$ are incorporated in the local variables $\gamma^\pm_N(D)$ and $M^\pm_P(D)$
and the dynamical effects from the Hamiltonian and finite momentum are considered in $M^\pm_P(D)$.

\section{Summary}\label{sec:summary}
A microscopic $n\alpha$ model was applied to $\Be8$, $\C12$, and $\O16$ systems
to describe the radial cluster motion in the ground and excited states. 
The positive- and negative-parity states were calculated with the GCM using the
generator coordinate $D$ for the $\alpha$-$\alpha$ distance.
The cluster motion in the coordinate $D$ space was analyzed, and the $\C12(+)_2$ and $\O16(+)_2$
states were found to be large-amplitude modes of radial excitation built on the ground states.

To describe the cluster motion of the $n\alpha$ systems, 
we proposed a collective model in the one-dimensional coordinate $D$
by utilizing inputs from the parity-projected microscopic $n\alpha$ wave functions.
The potential term in the collective Hamiltonian
was given by the energy expectation values of the $n\alpha$ wave function at $D$.
For the kinetic term in the collective Hamiltonian, a couple of prescriptions were tested.
To take into account the antisymmetrization effects on the coordinate space $D$,
the metric $\gamma_N$ derived from the norm kernel was considered.
To consider the dynamical effect,
we introduced an
imaginary shift $D\to D+iP/(2\hbar\nu)$ of the real parameter $D$ and defined
$\ket{D,P}$, in which
$D$ and $P$ represent the coordinate and momentum of the inter-cluster motion 
in the asymptotic region.
The mass $M_P$ was evaluated from the energy expectation value of $\ket{D,P}$, 
and was utilized to incorporate the dynamical effect on the collective mass of the kinetic
term in the collective Hamiltonian.

The collective wave functions of $n\alpha$ systems
were obtained by solving the collective model. The results of five sets of
metric $\gamma$ and mass $\mu$ in the collective model were compared with the
GCM results.
Among the five cases, the set $\gamma=1$ and $\mu=\gamma_N M_P$
of cal(2) was found to best reproduce the GCM results of the energy spectra, radii, and amplitude functions.
This corresponds to the prescription of the collective model described in
Eq.~\eqref{eq:coll-cal2}, in which the microscopic effects such as
antisymmetrization and parity projection are incorporated in the
parity- and $D$-dependent potential term and collective mass of the kinetic term.

\section*{Acknowledgments}

%\begin{acknowledgments}
%\blue{追加してください}
This work was supported
by Grants-in-Aid of the Japan Society for the Promotion of Science (Grant Nos. JP18K03617, JP18H05407, JP22K03633, and JP20K03964).
Discussions during the YIPQS international workshop
on ``Mean-field and Cluster Dynamics in Nuclear Systems" were useful to complete this work.

%\end{acknowledgments}

\appendix
\section{Physical coordinates} \label{app:coordinates}
In the AMD framework~\cite{Ono:1992uy}, the wave function of $A$-nucleon system is written by a Slater determinant
$\Phi_\textrm{AMD}=\textrm{det}[\psi_i(j)]$, where the single-nucleon wave function 
$\psi_i(j)=\varphi_{\bvec{Z}_i}(\bvec{r}_j) {\cal X}_i(\chi_j)$ is a product of the spatial wave function and 
the spin-isospin function ${\cal X}_i=\{p\uparrow, p\downarrow, n\uparrow, n\downarrow\}$. 
$\varphi_{\bvec{Z}}$ is given by a coherent state of a harmonic oscillator
\begin{align}
\varphi_{\bvec{Z}}(\bvec{r})=
\left(\frac{2\nu}{\pi}\right)^{4/3}
\exp\Bigl[-\nu({\bvec{r}}-\bvec{Z}/\sqrt{\nu})^2+\frac{1}{2} \bvec{Z}^2 \Bigr].
\end{align}
For the single-nucleon wave function, the mean position $\braket{\bvec{r}}$ and momentum
$\braket{\bvec{p}}$ are given by the real and imaginary parts of $\bvec{Z}$ as
\begin{align}
\braket{\bvec{r}}=\bvec{d},\quad \braket{\bvec{p}}=\bvec{k}, \\
\bvec{Z}=\sqrt{\nu}\bvec{d}+\frac{i}{2\hbar\sqrt{\nu}}\bvec{k}.
\end{align}
However, in the $A$-nucleon wave function, $\bvec{d}_{i}$ and $\bvec{k}_{i}$
indicate  positions and momenta of nucleons no longer 
because of the antisymmetrization.
Ono {\it et al.}~\cite{Ono:1992uy} introduced the physical coordinates $\bvec{W}_i$
instead of $\bvec{Z}_i$ as
\begin{align}
\bvec{W}_i \equiv \sum^A_{j=1} (\sqrt{Q})_{ij} \bvec{Z}_j,
\end{align}
where
\begin{align}
&Q_{ij} = B_{ij} B^{-1}_{ji} =
\frac{\partial}{\partial (\bvec{Z}^*_i\cdot \bvec{Z}_j )} \textrm{log} \bratoket{\Phi_\textrm{AMD}}{\Phi_\textrm{AMD}},
\\
&B_{ij}\equiv \bratoket{\psi_i}{\psi_j}=e^{\bvec{Z}^*_i\cdot \bvec{Z}_j} \bratoket{{\cal X}_i}{{\cal X}_j}, \\
&\bratoket{\Phi_\textrm{AMD}}{\Phi_\textrm{AMD}}=\textrm{det} B.
\end{align}
This is an extension of the physical coordinates in the $2\alpha$ system proposed by Saraceno
{\it et al.} in Ref.~\cite{Saraceno:1983hqo}.

For the present model space of the $2\alpha$, $3\alpha$, and $4\alpha$ systems,
$\bvec{Z}_i$ is taken to be $\bvec{Z}_i=\bvec{S}_m/\sqrt{\nu}$ $(i\in \alpha_m)$
with the real parameter $\bvec{S}_m$. Because of the symmetry of
spatial configurations of $\bvec{S}_m$, the physical coordinates are simply given as
$\bvec{W}_i=\lambda_{n\alpha}(D) \bvec{Z}_i$ with the scaling factors,
\begin{align}
&\lambda_{2\alpha}(D)=\sqrt{\frac{1+e^{-\nu D^2}}
{1-e^{-\nu D^2} } },\\
&\lambda_{3\alpha}(D)= \sqrt{\frac{1+e^{-\nu D^2/{2}}+e^{-\nu D^2}}
{1+e^{-\nu D^2/{2}}-2e^{-\nu D^2}} },\\
&\lambda_{4\alpha}(D)= \sqrt{\frac{1+2e^{-\nu D^2/{2}}-3e^{-\nu D^2}}
{1+e^{-\nu D^2/{2}}-2e^{-\nu D^2}} }.
\end{align}

\bibliography{na-refs}

\begin{thebibliography}{20}%
\makeatletter
\providecommand \@ifxundefined [1]{%
 \@ifx{#1\undefined}
}%
\providecommand \@ifnum [1]{%
 \ifnum #1\expandafter \@firstoftwo
 \else \expandafter \@secondoftwo
 \fi
}%
\providecommand \@ifx [1]{%
 \ifx #1\expandafter \@firstoftwo
 \else \expandafter \@secondoftwo
 \fi
}%
\providecommand \natexlab [1]{#1}%
\providecommand \enquote  [1]{``#1''}%
\providecommand \bibnamefont  [1]{#1}%
\providecommand \bibfnamefont [1]{#1}%
\providecommand \citenamefont [1]{#1}%
\providecommand \href@noop [0]{\@secondoftwo}%
\providecommand \href [0]{\begingroup \@sanitize@url \@href}%
\providecommand \@href[1]{\@@startlink{#1}\@@href}%
\providecommand \@@href[1]{\endgroup#1\@@endlink}%
\providecommand \@sanitize@url [0]{\catcode `\\12\catcode `\$12\catcode
  `\&12\catcode `\#12\catcode `\^12\catcode `\_12\catcode `\%12\relax}%
\providecommand \@@startlink[1]{}%
\providecommand \@@endlink[0]{}%
\providecommand \url  [0]{\begingroup\@sanitize@url \@url }%
\providecommand \@url [1]{\endgroup\@href {#1}{\urlprefix }}%
\providecommand \urlprefix  [0]{URL }%
\providecommand \Eprint [0]{\href }%
\providecommand \doibase [0]{https://doi.org/}%
\providecommand \selectlanguage [0]{\@gobble}%
\providecommand \bibinfo  [0]{\@secondoftwo}%
\providecommand \bibfield  [0]{\@secondoftwo}%
\providecommand \translation [1]{[#1]}%
\providecommand \BibitemOpen [0]{}%
\providecommand \bibitemStop [0]{}%
\providecommand \bibitemNoStop [0]{.\EOS\space}%
\providecommand \EOS [0]{\spacefactor3000\relax}%
\providecommand \BibitemShut  [1]{\csname bibitem#1\endcsname}%
\let\auto@bib@innerbib\@empty
%</preamble>
\bibitem [{\citenamefont {Hill}\ and\ \citenamefont
  {Wheeler}(1953)}]{Hill:1952jb}%
  \BibitemOpen
  \bibfield  {author} {\bibinfo {author} {\bibfnamefont {D.~L.}\ \bibnamefont
  {Hill}}\ and\ \bibinfo {author} {\bibfnamefont {J.~A.}\ \bibnamefont
  {Wheeler}},\ }\bibfield  {title} {\bibinfo {title} {{Nuclear Constitution and
  the Interpretation of Fission Phenomena}},\ }\href
  {https://doi.org/10.1103/PhysRev.89.1102} {\bibfield  {journal} {\bibinfo
  {journal} {Phys. Rev.}\ }\textbf {\bibinfo {volume} {89}},\ \bibinfo {pages}
  {1102} (\bibinfo {year} {1953})}\BibitemShut {NoStop}%
%%CITATION = PHRVA,89,1102;%%
\bibitem [{\citenamefont {Griffin}\ and\ \citenamefont
  {Wheeler}(1957)}]{Griffin:1957zza}%
  \BibitemOpen
  \bibfield  {author} {\bibinfo {author} {\bibfnamefont {J.~J.}\ \bibnamefont
  {Griffin}}\ and\ \bibinfo {author} {\bibfnamefont {J.~A.}\ \bibnamefont
  {Wheeler}},\ }\bibfield  {title} {\bibinfo {title} {{Collective Motions in
  Nuclei by the Method of Generator Coordinates}},\ }\href
  {https://doi.org/10.1103/PhysRev.108.311} {\bibfield  {journal} {\bibinfo
  {journal} {Phys. Rev.}\ }\textbf {\bibinfo {volume} {108}},\ \bibinfo {pages}
  {311} (\bibinfo {year} {1957})}\BibitemShut {NoStop}%
%%CITATION = PHRVA,108,311;%%
\bibitem [{\citenamefont {Brink}(1965)}]{Brink:1966}%
  \BibitemOpen
  \bibfield  {author} {\bibinfo {author} {\bibfnamefont {D.~M.}\ \bibnamefont
  {Brink}},\ }\bibfield  {title} {\bibinfo {title} {Many-body description of
  nuclear structure and reactions},\ }in\ \href@noop {} {\emph {\bibinfo
  {booktitle} {Proceedings of the International School of Physics ``E.
  Fermi"}}},\ Vol.\ \bibinfo {volume} {XXXV\hspace{-.1em}I},\ \bibinfo {editor}
  {edited by\ \bibinfo {editor} {\bibfnamefont {C.}~\bibnamefont {Bloch}}}\
  (\bibinfo  {publisher} {Academic, New York},\ \bibinfo {year} {1965})\ p.\
  \bibinfo {pages} {247}\BibitemShut {NoStop}%
\bibitem [{\citenamefont {Brink}\ and\ \citenamefont
  {Weiguny}(1968)}]{Brink:1968ybn}%
  \BibitemOpen
  \bibfield  {author} {\bibinfo {author} {\bibfnamefont {D.~M.}\ \bibnamefont
  {Brink}}\ and\ \bibinfo {author} {\bibfnamefont {A.}~\bibnamefont
  {Weiguny}},\ }\bibfield  {title} {\bibinfo {title} {{The generator coordinate
  theory of collective motion}},\ }\href
  {https://doi.org/10.1016/0375-9474(68)90059-6} {\bibfield  {journal}
  {\bibinfo  {journal} {Nucl. Phys. A}\ }\textbf {\bibinfo {volume} {120}},\
  \bibinfo {pages} {59} (\bibinfo {year} {1968})}\BibitemShut {NoStop}%
\bibitem [{\citenamefont {Horiuchi}(1970)}]{Horiuchi:1970}%
  \BibitemOpen
  \bibfield  {author} {\bibinfo {author} {\bibfnamefont {H.}~\bibnamefont
  {Horiuchi}},\ }\bibfield  {title} {\bibinfo {title} {{Generator Coordinate
  Treatment of Composite Particle Reaction and Molecule-like Structures}},\
  }\href {https://doi.org/10.1143/PTP.43.375} {\bibfield  {journal} {\bibinfo
  {journal} {Prog. Theor. Phys.}\ }\textbf {\bibinfo {volume} {43}},\ \bibinfo
  {pages} {375} (\bibinfo {year} {1970})}\BibitemShut {NoStop}%
\bibitem [{\citenamefont {Uegaki}\ \emph {et~al.}(1977)\citenamefont {Uegaki},
  \citenamefont {Okabe}, \citenamefont {Abe},\ and\ \citenamefont
  {Tanaka}}]{uegaki1}%
  \BibitemOpen
  \bibfield  {author} {\bibinfo {author} {\bibfnamefont {E.}~\bibnamefont
  {Uegaki}}, \bibinfo {author} {\bibfnamefont {S.}~\bibnamefont {Okabe}},
  \bibinfo {author} {\bibfnamefont {Y.}~\bibnamefont {Abe}},\ and\ \bibinfo
  {author} {\bibfnamefont {H.}~\bibnamefont {Tanaka}},\ }\bibfield  {title}
  {\bibinfo {title} {{Structure of the Excited States in $^{12}$C. I}},\ }\href
  {https://doi.org/10.1143/PTP.57.1262} {\bibfield  {journal} {\bibinfo
  {journal} {Prog. Theor. Phys.}\ }\textbf {\bibinfo {volume} {57}},\ \bibinfo
  {pages} {1262} (\bibinfo {year} {1977})}\BibitemShut {NoStop}%
\bibitem [{\citenamefont {Fujiwara}\ \emph {et~al.}(1980)\citenamefont
  {Fujiwara} \emph {et~al.}}]{Fujiwara-supp}%
  \BibitemOpen
  \bibfield  {author} {\bibinfo {author} {\bibfnamefont {Y.}~\bibnamefont
  {Fujiwara}} \emph {et~al.},\ }\bibfield  {title} {\bibinfo {title} {{Chapter
  II. Comprehensive Study of Alpha-Nuclei}},\ }\href
  {https://doi.org/10.1143/PTPS.68.29} {\bibfield  {journal} {\bibinfo
  {journal} {Prog. Theor. Phys. Suppl.}\ }\textbf {\bibinfo {volume} {68}},\
  \bibinfo {pages} {29} (\bibinfo {year} {1980})}\BibitemShut {NoStop}%
\bibitem [{\citenamefont {Descouvemont}\ and\ \citenamefont
  {Baye}(1987)}]{Descouvemont:1987zzb}%
  \BibitemOpen
  \bibfield  {author} {\bibinfo {author} {\bibfnamefont {P.}~\bibnamefont
  {Descouvemont}}\ and\ \bibinfo {author} {\bibfnamefont {D.}~\bibnamefont
  {Baye}},\ }\bibfield  {title} {\bibinfo {title} {{Microscopic theory of the
  $^{8}\mathrm{Be}$(\ensuremath{\alpha},\ensuremath{\gamma}${)}^{12}$C reaction
  in a three-cluster model}},\ }\href {https://doi.org/10.1103/PhysRevC.36.54}
  {\bibfield  {journal} {\bibinfo  {journal} {Phys. Rev. C}\ }\textbf {\bibinfo
  {volume} {36}},\ \bibinfo {pages} {54} (\bibinfo {year} {1987})}\BibitemShut
  {NoStop}%
%%CITATION = PHRVA,C36,54;%%
\bibitem [{\citenamefont {Libert-Heinemann}\ \emph {et~al.}(1980)\citenamefont
  {Libert-Heinemann}, \citenamefont {Baye},\ and\ \citenamefont
  {Heenen}}]{LIBERTHEINEMANN1980429}%
  \BibitemOpen
  \bibfield  {author} {\bibinfo {author} {\bibfnamefont {M.}~\bibnamefont
  {Libert-Heinemann}}, \bibinfo {author} {\bibfnamefont {D.}~\bibnamefont
  {Baye}},\ and\ \bibinfo {author} {\bibfnamefont {P.-H.}\ \bibnamefont
  {Heenen}},\ }\bibfield  {title} {\bibinfo {title} {Generator-coordinate study
  of inelastic $\alpha$+$^{12}\textrm{C}$ scattering},\ }\href
  {https://doi.org/https://doi.org/10.1016/0375-9474(80)90025-1} {\bibfield
  {journal} {\bibinfo  {journal} {Nucl. Phys. A}\ }\textbf {\bibinfo {volume}
  {339}},\ \bibinfo {pages} {429} (\bibinfo {year} {1980})}\BibitemShut
  {NoStop}%
\bibitem [{\citenamefont {Saito}(1969)}]{Saito:1969zz}%
  \BibitemOpen
  \bibfield  {author} {\bibinfo {author} {\bibfnamefont {S.}~\bibnamefont
  {Saito}},\ }\bibfield  {title} {\bibinfo {title} {{Interaction between
  Clusters and Pauli Principle}},\ }\href {https://doi.org/10.1143/PTP.41.705}
  {\bibfield  {journal} {\bibinfo  {journal} {Prog. Theor. Phys.}\ }\textbf
  {\bibinfo {volume} {41}},\ \bibinfo {pages} {705} (\bibinfo {year}
  {1969})}\BibitemShut {NoStop}%
\bibitem [{\citenamefont {Reinhard}\ and\ \citenamefont
  {Goeke}(1987)}]{Reinhard87}%
  \BibitemOpen
  \bibfield  {author} {\bibinfo {author} {\bibfnamefont {P.~G.}\ \bibnamefont
  {Reinhard}}\ and\ \bibinfo {author} {\bibfnamefont {K.}~\bibnamefont
  {Goeke}},\ }\bibfield  {title} {\bibinfo {title} {{The generator coordinate
  method and quantised collective motion in nuclear systems}},\ }\href
  {https://doi.org/10.1088/0034-4885/50/1/001} {\bibfield  {journal} {\bibinfo
  {journal} {Rep. Prog. Phys.}\ }\textbf {\bibinfo {volume} {50}},\ \bibinfo
  {pages} {1} (\bibinfo {year} {1987})}\BibitemShut {NoStop}%
\bibitem [{\citenamefont {Marumori}\ \emph {et~al.}(1980)\citenamefont
  {Marumori}, \citenamefont {Maskawa}, \citenamefont {Sakata},\ and\
  \citenamefont {Kuriyama}}]{Marumori:1980bu}%
  \BibitemOpen
  \bibfield  {author} {\bibinfo {author} {\bibfnamefont {T.}~\bibnamefont
  {Marumori}}, \bibinfo {author} {\bibfnamefont {T.}~\bibnamefont {Maskawa}},
  \bibinfo {author} {\bibfnamefont {F.}~\bibnamefont {Sakata}},\ and\ \bibinfo
  {author} {\bibfnamefont {A.}~\bibnamefont {Kuriyama}},\ }\bibfield  {title}
  {\bibinfo {title} {{Self-Consistent Collective-Coordinate Method for the
  Large-Amplitude Nuclear Collective Motion}},\ }\href
  {https://doi.org/10.1143/PTP.64.1294} {\bibfield  {journal} {\bibinfo
  {journal} {Prog. Theor. Phys.}\ }\textbf {\bibinfo {volume} {64}},\ \bibinfo
  {pages} {1294} (\bibinfo {year} {1980})}\BibitemShut {NoStop}%
\bibitem [{\citenamefont {Matsuo}\ \emph {et~al.}(2000)\citenamefont {Matsuo},
  \citenamefont {Nakatsukasa},\ and\ \citenamefont
  {Matsuyanagi}}]{Matsuo:2000hy}%
  \BibitemOpen
  \bibfield  {author} {\bibinfo {author} {\bibfnamefont {M.}~\bibnamefont
  {Matsuo}}, \bibinfo {author} {\bibfnamefont {T.}~\bibnamefont
  {Nakatsukasa}},\ and\ \bibinfo {author} {\bibfnamefont {K.}~\bibnamefont
  {Matsuyanagi}},\ }\bibfield  {title} {\bibinfo {title} {{Adiabatic
  Selfconsistent Collective Coordinate Method for Large Amplitude Collective
  Motion in Nuclei with Pairing Correlations}},\ }\href
  {https://doi.org/10.1143/PTP.103.959} {\bibfield  {journal} {\bibinfo
  {journal} {Prog. Theor. Phys.}\ }\textbf {\bibinfo {volume} {103}},\ \bibinfo
  {pages} {959} (\bibinfo {year} {2000})}\BibitemShut {NoStop}%
\bibitem [{\citenamefont {Hinohara}\ \emph {et~al.}(2008)\citenamefont
  {Hinohara}, \citenamefont {Nakatsukasa}, \citenamefont {Matsuo},\ and\
  \citenamefont {Matsuyanagi}}]{Hinohara:2007tj}%
  \BibitemOpen
  \bibfield  {author} {\bibinfo {author} {\bibfnamefont {N.}~\bibnamefont
  {Hinohara}}, \bibinfo {author} {\bibfnamefont {T.}~\bibnamefont
  {Nakatsukasa}}, \bibinfo {author} {\bibfnamefont {M.}~\bibnamefont
  {Matsuo}},\ and\ \bibinfo {author} {\bibfnamefont {K.}~\bibnamefont
  {Matsuyanagi}},\ }\bibfield  {title} {\bibinfo {title} {{Microscopic
  Derivation of Collective Hamiltonian by Means of the Adiabatic
  Self-Consistent Collective Coordinate Method: Shape Mixing in Low-Lying
  States of $^{68}$Se and $^{72}$Kr}},\ }\href
  {https://doi.org/10.1143/PTP.119.59} {\bibfield  {journal} {\bibinfo
  {journal} {Prog. Theor. Phys.}\ }\textbf {\bibinfo {volume} {119}},\ \bibinfo
  {pages} {59} (\bibinfo {year} {2008})}\BibitemShut {NoStop}%
\bibitem [{\citenamefont {Wen}\ and\ \citenamefont
  {Nakatsukasa}(2016)}]{Wen:2016buw}%
  \BibitemOpen
  \bibfield  {author} {\bibinfo {author} {\bibfnamefont {K.}~\bibnamefont
  {Wen}}\ and\ \bibinfo {author} {\bibfnamefont {T.}~\bibnamefont
  {Nakatsukasa}},\ }\bibfield  {title} {\bibinfo {title} {{Self-consistent
  collective coordinate for reaction path and inertial mass}},\ }\href
  {https://doi.org/10.1103/PhysRevC.94.054618} {\bibfield  {journal} {\bibinfo
  {journal} {Phys. Rev. C}\ }\textbf {\bibinfo {volume} {94}},\ \bibinfo
  {pages} {054618} (\bibinfo {year} {2016})}\BibitemShut {NoStop}%
\bibitem [{\citenamefont {Brink}\ \emph {et~al.}(1970)\citenamefont {Brink},
  \citenamefont {Friedrich}, \citenamefont {Weiguny},\ and\ \citenamefont
  {Wong}}]{Brink:1970ufk}%
  \BibitemOpen
  \bibfield  {author} {\bibinfo {author} {\bibfnamefont {D.~M.}\ \bibnamefont
  {Brink}}, \bibinfo {author} {\bibfnamefont {H.}~\bibnamefont {Friedrich}},
  \bibinfo {author} {\bibfnamefont {A.}~\bibnamefont {Weiguny}},\ and\ \bibinfo
  {author} {\bibfnamefont {C.~W.}\ \bibnamefont {Wong}},\ }\bibfield  {title}
  {\bibinfo {title} {{Investigation of the alpha-particle model for light
  nuclei}},\ }\href {https://doi.org/10.1016/0370-2693(70)90284-4} {\bibfield
  {journal} {\bibinfo  {journal} {Phys. Lett. B}\ }\textbf {\bibinfo {volume}
  {33}},\ \bibinfo {pages} {143} (\bibinfo {year} {1970})}\BibitemShut
  {NoStop}%
\bibitem [{\citenamefont {Volkov}(1965)}]{Volkov:1965zz}%
  \BibitemOpen
  \bibfield  {author} {\bibinfo {author} {\bibfnamefont {A.}~\bibnamefont
  {Volkov}},\ }\bibfield  {title} {\bibinfo {title} {{Equilibrium deformation
  calculations of the ground state energies of 1p shell nuclei}},\ }\href
  {https://doi.org/10.1016/0029-5582(65)90244-0} {\bibfield  {journal}
  {\bibinfo  {journal} {Nucl. Phys.}\ }\textbf {\bibinfo {volume} {74}},\
  \bibinfo {pages} {33} (\bibinfo {year} {1965})}\BibitemShut {NoStop}%
\bibitem [{\citenamefont {Kanada-En'yo}\ \emph {et~al.}(2014)\citenamefont
  {Kanada-En'yo}, \citenamefont {Suhara},\ and\ \citenamefont
  {Taniguchi}}]{Kanada-Enyo:2014mri}%
  \BibitemOpen
  \bibfield  {author} {\bibinfo {author} {\bibfnamefont {Y.}~\bibnamefont
  {Kanada-En'yo}}, \bibinfo {author} {\bibfnamefont {T.}~\bibnamefont
  {Suhara}},\ and\ \bibinfo {author} {\bibfnamefont {Y.}~\bibnamefont
  {Taniguchi}},\ }\bibfield  {title} {\bibinfo {title} {{Approximation of
  reduced width amplitude and application to cluster decay width}},\ }\href
  {https://doi.org/10.1093/ptep/ptu095} {\bibfield  {journal} {\bibinfo
  {journal} {Prog. Theor. Exp. Phys.}\ }\textbf {\bibinfo {volume} {2014}},\
  \bibinfo {pages} {073D02} (\bibinfo {year} {2014})}\BibitemShut {NoStop}%
\bibitem [{\citenamefont {Ono}\ \emph {et~al.}(1992)\citenamefont {Ono},
  \citenamefont {Horiuchi}, \citenamefont {Maruyama},\ and\ \citenamefont
  {Ohnishi}}]{Ono:1992uy}%
  \BibitemOpen
  \bibfield  {author} {\bibinfo {author} {\bibfnamefont {A.}~\bibnamefont
  {Ono}}, \bibinfo {author} {\bibfnamefont {H.}~\bibnamefont {Horiuchi}},
  \bibinfo {author} {\bibfnamefont {T.}~\bibnamefont {Maruyama}},\ and\
  \bibinfo {author} {\bibfnamefont {A.}~\bibnamefont {Ohnishi}},\ }\bibfield
  {title} {\bibinfo {title} {{Antisymmetrized Version of Molecular Dynamics
  with Two-Nucleon Collisions and Its Application to Heavy Ion Reactions}},\
  }\href {https://doi.org/10.1143/PTP.87.1185} {\bibfield  {journal} {\bibinfo
  {journal} {Prog. Theor. Phys.}\ }\textbf {\bibinfo {volume} {87}},\ \bibinfo
  {pages} {1185} (\bibinfo {year} {1992})}\BibitemShut {NoStop}%
\bibitem [{\citenamefont {Saraceno}\ \emph {et~al.}(1983)\citenamefont
  {Saraceno}, \citenamefont {Kramer},\ and\ \citenamefont
  {Fernandez}}]{Saraceno:1983hqo}%
  \BibitemOpen
  \bibfield  {author} {\bibinfo {author} {\bibfnamefont {M.}~\bibnamefont
  {Saraceno}}, \bibinfo {author} {\bibfnamefont {P.}~\bibnamefont {Kramer}},\
  and\ \bibinfo {author} {\bibfnamefont {F.}~\bibnamefont {Fernandez}},\
  }\bibfield  {title} {\bibinfo {title} {{Time-dependent variational
  description of \ensuremath{\alpha}\ensuremath{\alpha} scattering}},\ }\href
  {https://doi.org/10.1016/0375-9474(83)90325-1} {\bibfield  {journal}
  {\bibinfo  {journal} {Nucl. Phys. A}\ }\textbf {\bibinfo {volume} {405}},\
  \bibinfo {pages} {88} (\bibinfo {year} {1983})}\BibitemShut {NoStop}%
\end{thebibliography}%

\end{document}